\documentclass[aps,pre,twocolumn,showpacs,bibnotes,superscriptaddress,floatfix]{revtex4}

\usepackage{bibUrl}

\usepackage{amsmath}
\usepackage{amssymb}
\usepackage{amsthm}

\usepackage{graphicx}

\newcommand\pictc[5]{\begin{figure}
                   \centerline{
\includegraphics[width=#1\columnwidth,height=0.7\textheight,keepaspectratio]{#3}}
               \protect\caption{\protect\label{fig:#4} #5}
                \end{figure}            }
\newcommand\pict[4][1]{\pictc{#1}{!tb}{#2}{#3}{#4}}
\newcommand\rpict[1]{\ref{fig:#1}}

\newcommand\lsect[1]{\label{sect:#1}}
\newcommand\rsect[1]{\ref{sect:#1}}

\begin{document}
\begin{sloppy}

\title{Bifurcations and stability of gap solitons in periodic potentials}

\author{Dmitry E. Pelinovsky}
\affiliation{Department of Mathematics, McMaster University,
Hamilton, Ontario L8S 4K1, Canada} \affiliation{Nonlinear Physics
Group, Research School of Physical Sciences and Engineering,
Australian National University, Canberra, ACT 0200, Australia}
\homepage{dmpeli.math.mcmaster.ca}

\author{Andrey A. Sukhorukov}
\author{Yuri S. Kivshar}

\affiliation{Nonlinear Physics Group and Centre for Ultra-high
Bandwidth Devices for Optical Systems (CUDOS), Research School of
Physical Sciences and Engineering, Australian National University,
Canberra, ACT 0200, Australia}
\homepage{www.rsphysse.anu.edu.au/nonlinear}

\begin{abstract}
We analyze the existence, stability, and internal modes of {\em
gap solitons} in nonlinear periodic systems described by the
nonlinear Schr\"odinger equation with a sinusoidal potential, such
as photonic crystals, waveguide arrays, optically-induced photonic
lattices, and Bose-Einstein condensates loaded onto an optical
lattice. We study bifurcations of gap solitons from the band edges
of the Floquet-Bloch spectrum, and show that gap solitons can
appear near {\em all} lower or upper band edges of the spectrum,
for focusing or defocusing nonlinearity, respectively. We show
that, in general, {\em two types of gap solitons} can bifurcate
from each band edge, and one of those two is always {\em
unstable}. A gap soliton corresponding to a given band edge is
shown to possess a number of {\em internal modes} that bifurcate
from all band edges of the same polarity.
We demonstrate that stability of gap solitons is determined by
location of the internal modes with respect to the spectral bands
of the inverted spectrum and, when they overlap, complex
eigenvalues give rise to {\em oscillatory instabilities} of gap
solitons.
\end{abstract}


\maketitle

\section{Introduction} \lsect{intro}

Periodic structures are very common in physical problems, with the
crystalline lattice being the most familiar classical example. One
of the important features of such systems is the existence of {\em
multiple frequency gaps} in the wave transmission spectra. Such
spectral gaps are responsible for a strong modification of the
wave dispersion and diffraction that occurs when waves experience
resonant Bragg scattering from a periodic
structure~\cite{Yeh:1988:OpticalWaves}.

When nonlinear self-action becomes important, the systems with
periodically modulated parameters demonstrate a number of new
effects; in particular they can support a novel type of solitons,
the so-called {\em gap solitons}, which exist in the gaps of the
linear wave spectrum due to a strong Bragg scattering and coupling
between the forward and backward propagating
modes~\cite{Voloshchenko:1981-902:ZTF, Chen:1987-160:PRL, deSterke:1994-203:ProgressOptics}. During the last years,
it was shown that gap solitons may exist in different types of
nonlinear periodic structures including low-dimensional photonic
crystals and photonic layered structures~\cite{Akozbek:1998-2287:PRE, Mingaleev:2001-5474:PRL}, waveguide arrays~\cite{Mandelik:2003-53902:PRL, Mandelik:2004-93904:PRL}, optically-induced photonic lattices~\cite{Fleischer:2003-147:NAT, Neshev:nlin.PS/0311059:ARXIV}, and Bose-Einstein condensates loaded onto an optical lattice~\cite{Louis:2003-13602:PRA,
Ostrovskaya:2004-19:OE, Eiermann:cond-mat/0402178:ARXIV}.

There are known {\em two standard approaches} to study nonlinear
localized modes and gap solitons in periodic
structures~\cite{Kivshar:2003:OpticalSolitons}. The first approach
is based on the derivation of an effective {\em discrete nonlinear
Schr\"odinger equation} and the analysis of its stationary
localized solutions in the form of {\em discrete localized modes}
or {\em discrete solitons}~\cite{Christodoulides:1988-794:OL}. In the
solid-state physics, the similar approach is known as {\em the
tight-binding approximation} that, in application to periodic
photonic structures, corresponds to the case of weakly coupled
defect modes excited in each individual waveguide of the
structure. The analogous concepts appear in the study of other
systems such as the Bose-Einstein condensates in optical
lattices~\cite{Trombettoni:2001-2353:PRL}.

On the other hand, weak nonlinear effects in optical fibers with a
periodic modulation of the refractive index are well studied in
the framework of the other familiar and well-accepted approach,
{\em the coupled-mode theory}~\cite{deSterke:1994-203:ProgressOptics}. The coupled-mode theory for nonlinear periodic structures is based on a
decomposition of the wave field into the forward and backward
propagating modes, under the condition of the Bragg resonance, and
the derivation a system of coupled nonlinear equations
for those two modes. Such an approach is usually applied to
analyze nonlinear localized waves in the systems with a {\em
weakly modulated} optical refractive index known as gap (or Bragg)
solitons.

A number of recent experiments in the nonlinear guided wave optics~\cite{Mandelik:2003-53902:PRL, Christodoulides:2003-817:NAT, Sukhorukov:2004-93901:PRL, Mandelik:2004-93904:PRL, Neshev:nlin.PS/0311059:ARXIV} and Bose-Einstein condensates~\cite{Eiermann:2003-60402:PRL, Eiermann:cond-mat/0402178:ARXIV} were conducted in periodic structures 
under the conditions when none of those approximations is valid. Indeed, one of the main features of the wave propagation in periodic structures is the existence of a set of multiple forbidden gaps in the transmission spectrum. As a result, the nonlinearly-induced localization of waves can become possible in each of these gaps~\cite{deSterke:1994-203:ProgressOptics, Sukhorukov:2001-83901:PRL, Mandelik:2003-53902:PRL}. However, the effective discrete equation derived in the tight-binding approximation describes only {\em one transmission band} surrounded by {\em two semi-infinite band gaps} and, therefore, a fine structure of the band-gap spectrum associated with the wave transmission in periodic media
is lost in this approach. On the other hand, the
coupled-mode theory of gap solitons describes only 
{\em an isolated narrow gap} in between {\em two semi-infinite 
transmission bands}, and it does not allow to consider simultaneously the localized modes due to the total internal reflection as well as to study the band coupling and inter-band resonances. Recently, it was realized that the study of the simultaneous existence of localized modes of different types is a very important issue in the analysis of stability of nonlinear
localized modes and gap solitons~\cite{Sukhorukov:2001-83901:PRL, Sukhorukov:2003-2345:OL}.

The main purpose of this paper is to develop, for the first time
to our knowledge, a general analytical description of the existence, bifurcations, and stability of spatially localized nonlinear modes (i.e., lattice and gap solitons) in the framework of an effective continuous model
described by the nonlinear Schr\"odinger equation with a periodic
external potential. The use of this well-accepted nonlinear model
for our analysis allows us to remove all restrictions of both
approaches mentioned above, and to study consistently the effects
of the bandgap spectrum on the properties and linear stability of
gap solitons.

First, by applying the multi-scale asymptotic analytical methods,
we show that such gap solitons may appear in {\em all band gaps}
of the periodic potential for any sign of nonlinearity, but they
bifurcate from different band edges for different signs of
nonlinearity. Second, we demonstrate that, in general, only {\em
two branches of gap solitons} bifurcate from each band edge, and
one of those two is always linearly unstable. Third, we study
stability of gap solitons in a selected band gap and find the
soliton internal modes\cite{Kivshar:1998-5032:PRL, Pelinovsky:1998-121:PD} bifurcating from all other band edges of the same polarity. However, only
one internal mode can bifurcate from the band edge where the gap
soliton originates itself. At last, we analyze the conditions when
the bifurcation of the internal modes can give rise to complex
eigenvalues, which are shown to be responsible for {\em
oscillatory instabilities} of gap solitons due to single-gap~\cite{Barashenkov:1998-5117:PRL} or inter-gap~\cite{Sukhorukov:2001-83901:PRL, Sukhorukov:2003-2345:OL} resonances.

The paper is organized as follows. Section~\rsect{model} presents
our physical model which is described by an effective nonlinear
Schr\"{o}dinger equation with an external periodic potential of
the simplest sinusoidal form. Section~\rsect{bands} summarizes the
studies of the spectral properties of the linear eigenvalue
problem with a periodic potential. In 
Section~\rsect{bifurc} we study bifurcations of gap solitons by
means of the weakly nonlinear approximation.
Section~\rsect{branches} presents the analysis of the
exponentially small corrections beyond the weakly nonlinear
approximation. Section~\rsect{stability} discusses the stability
problem of gap solitons. Symmetry-breaking instabilities are
studied in Section~\rsect{splitting}. Internal and oscillatory
instability modes of gap solitons associated with non-zero
eigenvalues are studied in Section~\rsect{internal_mode}. Finally,
Sec.~\rsect{concl} summarizes our results and discuss further
perspectives. Appendix \rsect{Evans} 
gives details of the numerical method for calculations of eigenvalues.
Appendices~\rsect{appA} and~\rsect{appB} present
details of derivations, which are 
used in Sections~\rsect{splitting} and
~\rsect{internal_mode}, respectively.

\section{Model}  \lsect{model}

We consider the cubic nonlinear Schr\"{o}dinger (NLS)
equation with an external periodic potential in the form,
\begin{equation} \label{NLS}
     i \Psi_t = - \Psi_{xx} + V(x) \Psi + \sigma |\Psi|^2 \Psi,
\end{equation}
where $V(x + d) = V(x)$, $d$ is the fundamental period,
and $\sigma = \pm 1$ defines the type
of the wave self-action effect, namely {\em self-focusing}
($\sigma=-1$) or {\em self-defocusing} ($\sigma=+1$).
The analytical results presented below are rather general, and
they are valid for different types of smooth arbitrary-shaped
periodic potentials. However, in the numerical examples discussed
below we consider the squared sine potential,
\begin{equation} \label{sine}
   V(x) = V_0 \sin^2\left(\frac{\pi x}{d}\right).
\end{equation}
The harmonic potential (\ref{sine}) describes, in the mean-field
approximation, the dynamics of the Bose-Einstein condensate in
an optical lattice, when the parabolic trap is
neglected~\cite{Louis:2003-13602:PRA,Ostrovskaya:2004-19:OE}.
The squared sine potential $V(x)$ has two extremum points on
the period of $x$, such that $x = 0$ is the minimum
and $x = d/2$ is the maximum of $V(x)$.

Stationary localized solutions of the cubic NLS equation
(\ref{NLS}) for gap solitons are found in the form $\Psi(x,t) =
\psi(x) \exp(-i \mu t)$, where $\mu$ is referred to as {\em the
soliton parameter}. The soliton profile $\psi(x)$ is found as a
spatially localized solution of the nonlinear problem:
\begin{equation} \label{elliptic}
   - \psi^{\prime \prime} + V(x) \psi + \sigma |\psi|^2 \psi = \mu
   \psi,
\end{equation}
where the prime stands for a derivative in $x$. Existence and
multiplicity of multi-humped localized states $\psi(x)$ in the
spectral gaps of the periodic potential $V(x)$ were considered by
means of the variational methods by Alama and
Li~\cite{Alama:1992-89:JDE, Alama:1992-983:IUMJ}. Bifurcations of
bound states were analyzed by Kupper and
Stuart~\cite{Kupper:1990-1:JRAM, Kupper:1992-893:NATM} and Heinz
and Stuart~\cite{Heinz:1992-145:NATM} who proved that, depending
on the sign of the nonlinear term, lower or upper end-points of
the continuous spectrum are bifurcation points. Extension to the
multi-dimensional case was developed with Bloch waves of the
linear Schr\"{o}dinger operator~\cite{Heinz:1992-341:JDE}.
The number of branches of bound states was classified in terms of
the eigenvalues of the linear Schr\"{o}dinger operator with
periodic and decaying potentials~\cite{Heinz:1995-149:JDE}.
Eigenvalues of the latter (linear) problem were previously
considered by Gesztesy {\em et al.}~\cite{Gesztesy:1988-597:CMP}
and Alama {\em et al.}~\cite{Alama:1989-291:CMP}. In application
to the problem of the Bose-Einstein condensates in optical
lattices, the stationary model (\ref{elliptic}) has been considered
recently by Louis {\em et al.}~\cite{Louis:2003-13602:PRA} who found numerically
different types of spatially localized solutions in different band gaps.

All previous results were restricted to the study of the existence
of spatially localized solutions. Here, we use more general
methods (but, in some sense, less rigorous from the mathematical
point of view) and study bifurcations, stability, and internal
modes of gap solitons. To achieve these objectives, we apply
the multi-scale perturbation series expansion methods, developed
earlier by Iizuka~\cite{Iizuka:1994-4343:JPSJ} and Iizuka and
Wadati~\cite{Iizuka:1997-2308:JPSJ}. With the perturbation series
methods, we classify systematically the branches of gap solitons
bifurcating from the band edges to the band gaps, as well as their
stability.

\section{Spectral bands and gaps}   \lsect{bands}

Periodic potential $V(x)$ induces a band-gap structure in the
linear Schr\"{o}dinger spectral problem:
\begin{equation} \label{spectrum}
   - \psi'' + V(x) \psi = \mu \psi.
\end{equation}
The spectral bands are located for $\mu \in \Sigma_{\rm band}$,
where we enumerate the band edges in the following order:
\begin{equation}
\Sigma_{\rm band} = [\mu_0,\mu_1] \cup [\mu_3,\mu_2] \cup
[\mu_4,\mu_5] \cup [\mu_7,\mu_6] ...
\end{equation}
The spectral bands are
computed for the squared sine potential (\ref{sine}) and the
results are shown on Figure~\rpict{bands}. Review of spectral
theory for periodic potentials can be found in the book by
Eastham~\cite{Eastham:1973:SpectralTheory}. Here we recover some
details which are important for our analysis. 

\pict{fig01.eps}{bands}{ The structure of spectral bands of the linear
periodic problem (\ref{spectrum}): (a) Trace of fundamental matrix
$\Delta$ versus $\mu$; (b) Floquet exponent $k$ versus $\mu$; 
(c) Solid: Bloch waves at the band
edges, as indicated by arrows; Dashed: potential $V(x)$.
Parameters are $V_0 = 1$ and $d=10$.}

When the spectral parameter $\mu$ is taken inside the spectral
bands, i.e. $\mu \in \Sigma_{\rm band}$, the problem
(\ref{spectrum}) has two linearly independent solutions in the form of Bloch waves,
\begin{equation} \label{fundamental}
   \psi_1 = \phi_1(x) e^{ik(\mu)x}, \qquad \psi_2
   = \phi_2(x) e^{-ik(\mu)x},
\end{equation}
where $\phi_{1,2}(x)$ are periodic functions and $k(\mu)$ is the
Floquet exponent, which can be chosen inside the first Brillouin
zone such that $0 \leq k(\mu) \leq \frac{\pi}{d}$. The graph of
$k(\mu)$ is shown in Fig.~\rpict{bands}(b) for the first three
spectral bands.

The spectral bands of the periodic potential (\ref{spectrum}) are
described by the function $\Delta(\mu)$, which is the trace of
fundamental matrix of solutions~\cite{Eastham:1973:SpectralTheory}
\begin{equation} \label{trace}
   \Delta(\mu) = 2 \cos k(\mu)d,
\end{equation}
The spectral bands are defined for $-2 \leq \Delta(\mu) \leq 2$,
which corresponds to propagating waves with real $k$. On the other
hand, the waves become exponentially localized inside the gaps,
where $|\Delta(\mu)| > 2$ and ${\rm Im}(k) \ne 0$. A
characteristic dependence $\Delta(\mu)$ is displayed in
Fig.~\rpict{bands}(a).

There are infinitely many spectral bands for a one-dimensional
periodic potential $V(x)$, where $|\Delta(\mu)| \leq
2$~\cite{Eastham:1973:SpectralTheory}. If $\Delta'(\mu_n) \neq 0$
at the band edge $\mu = \mu_n$, two adjacent spectral bands do not
overlap, such that the corresponding band gap has a non-zero
width. We consider the non-degenerate spectral band, such that
$\Delta'(\mu_n) \neq 0$ at the end point $\mu = \mu_n$.

The even-numbered band edges $\mu = \mu_{2m}$, $m \geq 0$
correspond to periodic Bloch functions $\psi_{2m}(x+d) =
\psi_{2m}(x)$, while the odd-numbered band edges $\mu =
\mu_{2m-1}$, $m \geq 1$ correspond to anti-periodic Bloch
functions $\psi_{2m-1}(x+d)=-\psi_{2m-1}(x)$. The Bloch functions
$\psi_n(x)$ for the first five band edges $\mu =
\mu_0,\mu_1,\mu_3,\mu_2,\mu_4$ are shown on Fig.~\rpict{bands}(c).

We now demonstrate that the bifurcations of bound states and
stationary gap solitons may occur when the two fundamental
solutions $\psi_{1,2}(x)$ in (\ref{fundamental}) become linearly
dependent. Since $\phi_1(x) e^{2 i k(\mu) x}$ solves the same
equation as $\phi_2(x)$ but it is not a periodic function of $x$,
unless $k(\mu) = 0 \;\left({\rm mod}\left(\frac{2
\pi}{d}\right)\right)$ or $k(\mu) = \frac{\pi}{d} \;\left({\rm
mod}\left(\frac{2 \pi}{d}\right)\right)$, the two solutions
$\psi_{1,2}(x)$ are always linearly independent in the interior of
the spectral bands $\mu \in \Sigma_{\rm band}$. 
On the other hand, the two solutions $\psi_{1,2}(x)$
become linearly dependent at the band edges $\mu = \mu_n$, since
\begin{equation}
   \psi_{2m}(x) = \phi_1(x) = \phi_2(x)
\end{equation}
and $k(\mu_{2m}) = 0 \;\left({\rm mod}\left(\frac{2 \pi}{d}\right)\right)$
at the even-numbered band edges, and
\begin{equation}
   \psi_{2m-1}(x) = \phi_1(x) e^{\frac{\pi i x}{d}}
   = \phi_2(x) e^{\frac{- \pi i x}{d}}
\end{equation}
and $k(\mu_{2m-1}) = \frac{\pi}{d} \;\left({\rm mod}\left(\frac{2 \pi}{d}\right)\right)$
at the odd-numbered band edges. The band edge $\mu = \mu_n$ has
geometric multiplicity one with the only linearly independent
Bloch function $\psi_n(x)$. The second, linearly independent
solution of (\ref{spectrum}) at $\mu = \mu_n$ grows linearly in
$x$. The band edge $\mu = \mu_n$ has however algebraic
multiplicity two, since there exists a generalized Bloch function
$\psi_{n}^{(1)}(x)$ that solves the derivative problem:
\begin{equation} \label{derivative}
   - (\psi_{n}^{(1)})'' + V(x) \psi_n^{(1)} -
   \mu_n \psi_n^{(1)} = 2 \psi_n'(x).
\end{equation}
It follows from (\ref{derivative}) that the generalized Bloch
functions $\psi_{2m}^{(1)}(x)$ and $\psi_{2m-1}^{(1)}(x)$ are
periodic and anti-periodic in $x$, respectively. We conclude that
the band edges $\mu = \mu_n$ are the only bifurcation points of
the linear spectrum $\mu \in \Sigma_{\rm band}$, associated with
the periodic potential $V(x)$.

The band curvature near the band edge $\mu = \mu_n$ follows from
the expansion of $\Delta(\mu)$ defined in (\ref{trace}):
\begin{eqnarray}
    \nonumber \Delta(\mu_n) + \Delta'(\mu_n) (\mu - \mu_n) +
   {\rm O}(\mu - \mu_n)^2 \\
    \label{expansions}
    = (-1)^n \left( 2 - d^2 (k - k_n)^2 + {\rm O}(k - k_n)^4 \right),
\end{eqnarray}
where $k_{2m} = 0$ and $k_{2m-1} = \frac{\pi}{d}$. As a result,
$\Delta(\mu_n) = 2 (-1)^n$ and
\begin{equation} \label{curvature}
   \mu = \mu_n - \mu_n^{(2)} ( k - k_n)^2 + {\rm
   O}(k - k_n)^4,
\end{equation}
such that
\begin{equation*}
   \mu_n^{(2)} = \frac{d^2 (-1)^n}{\Delta'(\mu_n)}.
\end{equation*}

The band curvatures $\mu_{n}^{(2)}$ can be expressed in terms of
Bloch functions $\psi_n(x)$ and $\psi_n^{(1)}(x)$ at $\mu =
\mu_n$. We use perturbation series expansions for fundamental
solutions $\phi_{1,2}(x)$ near the band edges $\mu = \mu_n$:
\begin{eqnarray} \nonumber
   \phi_{1,2}(x) e^{\pm i k_n x} =
   \psi_n(x) \pm i (k - k_n) \psi_n^{(1)}(x) \\
   \label{perturbation1} - (k - k_n)^2
   \psi_n^{(2)}(x) + {\rm O}(k - k_n)^3.
\end{eqnarray}
The eigenvalue $\mu$ is expanded in the perturbation series
(\ref{curvature}). The second-order correction $\psi_n^{(2)}(x)$
satisfies the non-homogeneous linear equation:
\begin{eqnarray} 
\nonumber
   - \left( \psi_n^{(2)} \right)'' + V(x)
   \psi_n^{(2)} - \mu_n \psi_n^{(2)} \\
\label{nonhomogeneous1}
    = 2 \left( \psi_n^{(1)}\right)'
   + \left( 1 + \mu_n^{(2)} \right) \psi_n.
\end{eqnarray}
If $\phi_{1,2}(x)$ are periodic functions of $x$, the second-order
correction $\psi_n^{(2)}(x)$ in the perturbation series
(\ref{perturbation1}) is periodic for $n = 2m$ and anti-periodic
for $n = 2m -1$. By Fredholm Alternative, this condition is
satisfied if the right-hand-side of (\ref{nonhomogeneous1})
satisfies the constraint:
\begin{equation} \label{Fredholm1}
   \left( 1 + \mu_n^{(2)} \right) \int_0^d \psi_n^2
   dx + 2 \int_0^d \psi_n \left( \psi_n^{(1)}\right)' dx = 0.
\end{equation}
Therefore, the band curvature $\mu_n^{(2)}$ is expressed in terms
of integrals of $\psi_n(x)$ and $\psi_n^{(1)}(x)$. The
perturbation series expansions (\ref{curvature}) and
(\ref{perturbation1}) can be continued algorithmically to the
higher orders in powers of $(k - k_n)$.

\section{Bifurcations of gap solitons} \lsect{bifurc}

Nonlinear bound states (gap solitons) of the NLS equation
(\ref{NLS}) are stationary solutions in the form:
\begin{equation} \label{soliton}
   \Psi(x,t) = \Phi_s(x) e^{-i\mu_s t},
\end{equation}
where the real function $\Phi_s(x)$ decays to zero as
$|x| \to \infty$ and satisfies the nonlinear
problem,
\begin{equation} \label{ODE}
   - \Phi_s'' + V(x) \Phi_s + \sigma \Phi_s^3 = \mu_s
   \Phi_s.
\end{equation}
When $\Phi_s(x)$ is small, the nonlinear potential $\sigma \Phi_s^2(x)$
acts as a perturbation term to the periodic potential $V(x)$. The
perturbation term leads to bifurcation of gap solitons $\Phi_s(x)$
from the band edges $\mu = \mu_n$ of the linear band-gap spectrum.
We study bifurcations of gap solitons with the multi-scale
perturbation series expansions:
\begin{equation} \label{perturbation2}
   \mu_s = \mu_n + \epsilon^2 \Delta_n + {\rm O}(\epsilon^4),
\end{equation}
and
\begin{equation} \label{soliton-form}
   \Phi_s(x) = \epsilon \Phi_{\epsilon}(x;X),
   \quad X = \epsilon(x-x_0), \quad \epsilon \ll 1,
\end{equation}
where
\begin{eqnarray} \nonumber
   \Phi_{\epsilon}(x;X) & = & A_n(X) \psi_n(x)
   + \epsilon A_n'(X) \psi_n^{(1)}(x) \\
   \label{perturbation3} & + & \epsilon^2 \phi_s^{(2)}(x;X)
   + {\rm O}(\epsilon^3).
\end{eqnarray}
Here $A_n(X)$ is a space-decaying bound state and $\psi_n(x)$ is
the periodic or anti-periodic Bloch function. Parameter $x_0$
determines a location of $A_n(X)$ with respect to $\psi_n(x)$. The
Bloch functions $\psi_n(x)$ and $\psi_n^{(1)}(x)$ are defined from
the linear problems (\ref{spectrum}) and (\ref{derivative}).

The second-order correction term $\phi_s^{(2)}(x;X)$ satisfies the
linear non-homogeneous equation:
\begin{eqnarray} \nonumber
  - \left(\phi_s^{(2)}\right)'' + V(x)
  \phi_s^{(2)} - \mu_n \phi_s^{(2)} \\
  \label{nonhomogeneous2} = A_n'' \left( \psi_n + 2
  \left(\psi_n^{(1)}\right)' \right) - \sigma A_n^3 \psi_n^3 +
  \Delta_n A_n \psi_n.
\end{eqnarray}
The secular growth of $\phi_s^{(2)}(x;X)$ in $x$ is removed if the
right-hand-side of (\ref{nonhomogeneous2}) satisfies the Fredholm
condition, which reduces to the nonlinear equation for $A_n =
A_n(X)$:
\begin{equation} \label{NLSapproximation}
   \mu_n^{(2)} A_n'' + \chi_n^{(2)} A_n^3 -
   \Delta_n A_n = 0,
\end{equation}
where
\begin{equation}
   \chi_n^{(2)} = \sigma \frac{\int_0^d \psi_n^4 dx}{ \int_0^d
   \psi_n^2 dx}.
\end{equation}
Using the constraint (\ref{NLSapproximation}), we represent the
second-order correction term $\phi_s^{(2)}(x;X)$ in the form:
\begin{equation}
   \phi_s^{(2)}(x;X) = A_n''(X) \psi_n^{(2)}(x) + A_n^3(X)
   \psi_n^{(nl2)}(x),
\end{equation}
where $\psi_n^{(2)}(x)$ solves the non-homogeneous problem
(\ref{nonhomogeneous1}), while $\psi_n^{(nl2)}(x)$ solves the
problem,
\begin{equation} \label{nonhomogeneous22}
   - \left(\psi_n^{(nl2)}\right)'' + V(x)
   \psi_n^{(nl2)} - \mu_n \psi_n^{(nl2)} = \chi_n^{(2)} \psi_n -
   \sigma \psi_n^3.
\end{equation}

The nonlinear equation (\ref{NLSapproximation}) is just the
stationary NLS equation, which has sech-solitons if ${\rm
sign}(\mu_n^{(2)}) = {\rm sign}(\chi_n^{(2)}) = {\rm
sign}(\Delta_n)$. For the focusing nonlinearity, $\sigma = -1$,
the sech-solitons bifurcate from all band edges, where
$\mu_n^{(2)} < 0$, such that $\Delta_n < 0$. It follows 
from (\ref{curvature}) 
that branches of gap solitons detach from all lower band edges
downwards the corresponding band gaps [see Fig.~\rpict{bands}(a)].
For the defocusing nonlinearity, $\sigma = +1$, the sech-solitons
bifurcate from all band edges, where $\mu_n^{(2)} > 0$, such that
$\Delta_n > 0$. Therefore, branches of gap solitons detach from
all upper band edges upwards the corresponding band gaps [see
Fig.~\rpict{bands}(b)]. Branches of gap solitons are shown on
Fig.~\rpict{solitonSF} for $\sigma = -1$ near the band edges $\mu
= \mu_0$, $\mu = \mu_3$, and $\mu = \mu_4$ and on
Fig.~\rpict{solitonDF} for $\sigma = +1$ near the band edges $\mu
= \mu_1$ and $\mu = \mu_2$. The families of gap solitons have been
found by solving the nonlinear eigenvalue problem (\ref{ODE}) with
a standard relaxation
technique~\cite{Press:1992:NumericalRecipes}.

\pict{fig02.eps}{solitonSF}{ Bifurcations for the on-site and
off-site gap solitons in a self-focusing medium ($\sigma=-1$).
Top: the soliton power $P(\mu) =
\int^{\infty}_{-\infty}\Phi_s^2(x;\mu) dx$ versus $\mu$. Solid:
solitons centered at $x_0=0$, dashed: centered at $x_0=d/2$.
(a-f): spatial profiles of gap solitons corresponding to marked
points in the upper plot; shading marks the minima of the
potential $V(x)$. }

The sech-solitons of the nonlinear equation 
(\ref{NLSapproximation}) are written explicitly in the form:
\begin{equation} \label{NLSsoliton}
   A_n(X) = a_n \; {\rm sech} (\kappa_n X),
\end{equation}
where $a_n$ and $\kappa_n$ are found from equations,
\begin{equation}
   \Delta_n - \kappa_n^2 \mu_n^{(2)} = 0, \qquad \chi_n^{(2)} a_n^2 -
   2 \kappa_n^2 \mu_n^{(2)} = 0,
\end{equation}
provided that ${\rm sign}(\mu_n^{(2)}) = {\rm sign}(\chi_n^{(2)})
= {\rm sign}(\Delta_n)$. The sech-type soliton envelopes
(\ref{NLSsoliton}) always have single-humped profile. Since
$A_n(X)$ is the envelope of $\psi_n(x)$, the resulting
nonlinear bound state $\Phi_s(x)$ has the oscillatory structure
near the band edge $\mu_s = \mu_n$.

\pict{fig03.eps}{solitonDF}{ Bifurcations for the on-site and
off-site gap solitons in a self-defocusing medium ($\sigma=+1$).
Notations are the same as on Figure~\rpict{solitonSF}. }

\section{Branches of gap solitons due to symmetry breaking}
        \lsect{branches}

The absence of translational invariance along the $x$ direction,
associated with the presence of the periodic potential, has an
important effect on the soliton properties. For example, it was
found that discrete solitons, bifurcating from the first band, can
be centered at (on-site) or in-between (off-site) potential wells.
In this Section, we demonstrate that two branches of gap solitons, 
bifurcating from all the bands, are centered at different 
positions in the periodic potential.

The gap soliton $\Phi_s(x)$ near the band edge $\mu_s = \mu_n$ is
represented by the perturbation series expansions
(\ref{perturbation2})--(\ref{perturbation3}), provided that the
formal series converges. Parameter $x_0$ in the "slow" coordinate
$X = \epsilon(x-x_0)$ determines the location of the bound state
$A_n(X)$ with respect to the Bloch function $\psi_n(x)$. We will
show that only two values of $x_0$ on the period of $x$
secure convergence of the formal series, in the general case. Our
analysis is equivalent to the construction of the Melnikov
function, which gives the distance between separatrices in the
nonlinear oscillator with a small, rapidly varying force
\cite{Gelfreich:1997-227:PD,Gelfreich:2000-266:PD}. Zeros of the
Melnikov function indicate values of $x_0$, where the separatrices
intersect, so that a homoclinic orbit for the gap soliton exists
in the nonlinear problem (\ref{ODE}) with the periodic potential
$V(x)$.

We will derive the Melnikov function
\cite{Gelfreich:1997-227:PD,Gelfreich:2000-266:PD} with a simple
but equivalent method. Derivative of the nonlinear equation
(\ref{ODE}) in $x$ results in the following third-order ODE:
\begin{equation}  \label{shift}
   - \Phi_s''' + V(x) \Phi_s' - \mu_s \Phi_s' + 3
   \sigma \Phi_s^2 \Phi_s' + V'(x) \Phi_s = 0.
\end{equation}
If the gap soliton $\Phi_s(x)$ exists, then it satisfies zero boundary
conditions as $|x| \to \infty$. Multiplication of
(\ref{shift}) by $\Phi_s(x)$ and integration it over $x$
result in the following constraint:
\begin{equation} \label{constraint}
   M_s(x_0) = \int_{-\infty}^{\infty} V'(x)
   \Phi_s^2(x) dx = 0.
\end{equation}
The function $M_s(x_0)$ is the Melnikov function
for existence of homoclinic orbits \cite{Gelfreich:1997-227:PD,Gelfreich:2000-266:PD}.
The constraint (\ref{constraint}) is always satisfied if the gap
soliton $\Phi_s(x)$ and the potential $V(x)$ are symmetric with
respect to the location of the central peak at $x = x_0$, such
that $\Phi_s^2(x-x_0) = \Phi_s^2(x_0-x)$ and $V'(x-x_0) =
-V'(x_0-x)$. More precise information on the constraint
(\ref{constraint}) can be obtained near the band edge $\mu_s =
\mu_n$, where the perturbation series expansions (\ref{perturbation2})--
(\ref{perturbation3}) are valid. The function $\Phi_{\epsilon}(x;X)$
has the power series expansion in $\epsilon$, each term of which
satisfies the squared-periodic boundary conditions in $x$,
\begin{equation} \label{per-van}
   \Phi^2_{\epsilon}(x+d;X) = \Phi^2_{\epsilon}(x;X),
\end{equation}
and the decaying boundary conditions in $X$,
\begin{equation} \label{van-per}
   \lim_{|X| \to \infty} \Phi_{\epsilon}(x;X) = 0.
\end{equation}
We shall prove that $M_s(x_0)$ is exponentially small in terms
of $\epsilon$. To do so, we rewrite equation (\ref{shift}) for
$\Phi_{\epsilon}(x;X)$,
multiply it by $\Phi_{\epsilon}(x;X)$ and integrate the resulting
equation over $x \in [0,d]$. Using the periodic boundary condition
(\ref{per-van}), we derive the relation,
\begin{eqnarray}
\nonumber
   \int_0^d V'(x) \Phi_{\epsilon}^2(x;X) dx \\
   = - 2 \epsilon \frac{\partial}{\partial X}
   \int_0^d \left( \Phi_{\epsilon, x}\right)^2 dx
   - 2 \epsilon^2 \frac{\partial}{\partial X}
   \int_0^d \Phi_{\epsilon, x} \Phi_{\epsilon, X} dx.
\end{eqnarray}
Using the decaying boundary condition (\ref{van-per}), we prove
that
\begin{equation} \label{constraint1}
   \int_{-\infty}^{\infty} dX \int_0^d V'(x)
   \Phi_{\epsilon}^2(x;X) dx = 0.
\end{equation}
As a result, the function $V'(x) \Phi_{\epsilon}^2(x;X)$ is
expanded in Fourier series in $x$ as
\begin{equation} \label{Fourier}
   V'(x) \Phi_{\epsilon}^2(x;X) = \sum_{m =
   -\infty}^{\infty} W_{n,m}(X;\epsilon) e^{\frac{2 \pi i m x}{d}},
\end{equation}
such that $W_{n,-m}(X;\epsilon) = \overline{W_{n,m}}(X;\epsilon)$
and
\begin{equation} \label{zero-term-van}
  \int_{-\infty}^{\infty} W_{n,0}(X;\epsilon)
  dX = 0
\end{equation}
at any order of $\epsilon$. The Fourier transform of $F(X)$ is
defined by the standard integral:
\begin{equation} \label{Fouriertransform}
   \hat{F}(k) = \int_{-\infty}^{\infty} F(X)
   e^{ikX} dX.
\end{equation}
The Melnikov function $M_s(x_0)$ is then expanded with the use of
the Fourier series (\ref{Fourier}) and the Fourier transform
(\ref{Fouriertransform}) in the form:
\begin{equation} \label{constraintsymplified}
   M_s(x_0) = \epsilon \sum_{m =
   -\infty}^{\infty} \hat{W}_{n,m} \left(\frac{2 \pi m}{\epsilon
   d};\epsilon\right) e^{\frac{2 \pi i m x_0}{d}}.
\end{equation}
At the leading order, we have $W_{n,m}(X;0) = A_n^2(X) w_{n,m}^{(0)}$,
where $w_{n,m}^{(0)}$ are coefficients in the Fourier series,
\begin{equation} \label{Fourier1}
   V'(x) \psi_n^2(x) = \sum_{m = -\infty}^{\infty}
   w_{n,m}^{(0)} e^{\frac{2 \pi i m x}{d}}.
\end{equation}
The zero-order term ($m=0$) in the series
(\ref{constraintsymplified}) is zero at any order of $\epsilon$,
since the constraint (\ref{zero-term-van}) results in the
condition: $\hat{W}_{n,0}(0;\epsilon) = 0$. The higher-order terms
with larger values of $|m|$ are exponentially smaller compared to
the terms with smaller values of $|m|$ in the limit $\epsilon \to
0$, since $\hat{A}_n^2(k)$ is exponentially decaying in $k$.
Therefore, using exponential asymptotics, we truncate the series
(\ref{constraintsymplified}) by the first-order terms ($m = \pm
1$) in the limit $\epsilon \to 0$:
\begin{equation} \label{roots-bif}
   M_s(x_0) = \epsilon \Lambda_1 \cos\left(\frac{2 \pi x_0}{d}
   + {\rm arg}(w_{n,1}^{(0)})\right) + E_1,
\end{equation}
where
\begin{equation*}
    \Lambda_1 = 2 |w_{n,1}^{(0)}| \hat{A}_n^2\left(\frac{2 \pi}{\epsilon d} \right)
\end{equation*}
and
\begin{equation*}
    E_1 = {\rm O}\left(\epsilon^2 \hat{A}_n^2\left(\frac{2 \pi}{\epsilon d} \right)\right)
    + {\rm O}\left(\epsilon \hat{A}_n^2\left(\frac{4 \pi}{\epsilon d} \right)\right).
\end{equation*}
Assuming that $\Lambda_1 \neq 0$, we conclude from
(\ref{roots-bif}) that there are precisely two families of gap
solitons bifurcating from two roots of the function
$\cos(\frac{2\pi x_0}{d})$ on the period of $x_0$.

We now prove that, for the squared sine potential~(\ref{sine}),
the values of $\arg(w_{n,1}^{(0)})$ are the same for all band edges
as $\arg(w_{n,1}^{(0)}) = - \frac{\pi}{2}$. It is clear from~(\ref{sine})
that $V(-x) = V(x)$ and $V'(x) \ge 0$ for $0 \le |x| \le d/2$,
while all Bloch wave squared amplitudes are symmetric, 
such that $\psi_n^2(-x) = \psi_n^2(x)$. As a result,
it follows from (\ref{Fourier1}) that 
$\arg(w_{n,1}^{(0)}) = - \frac{\pi}{2}$
and the two roots of $x_0$ occur at extremal points of $V(x)$:
$x_0 = 0$ and $x_0 = \frac{d}{2}$. The former (minimum)
point corresponds to the on-site gap soliton, while the latter
(maximum) point corresponds to the off-site gap soliton, in
accordance with Figures~\rpict{solitonSF} and~\rpict{solitonDF}.

When $\Lambda_1 = 0$ and $\hat{W}_{n,1}\left( \frac{2\pi}{\epsilon
d};\epsilon \right) \neq 0$, higher powers of $\epsilon$ are generally
non-zero in the first-order terms $(m = \pm 1)$, such that only
two branches of gap solitons $\Phi_s(x)$ bifurcate in a general
case. If the potential $V(x)$ is special such that
$\hat{W}_{n,1}\left(\frac{2 \pi}{\epsilon d};\epsilon\right) = 0$
at any order of $\epsilon$ but $\hat{W}_{n,2}\left(\frac{4
\pi}{\epsilon d};\epsilon\right) \neq 0$, the leading-order terms
in the series (\ref{constraintsymplified}) become second-order ($m
= \pm 2$), such that four branches of gap solitons $\Phi_s(x)$ may
bifurcate from four roots of the function $\cos(\frac{4\pi
x_0}{d})$ on the period of $x_0$. We do not know whether
any special potentials $V(x)$ may exist to hold the constraint
$\hat{W}_{n,1}\left(\frac{2 \pi}{\epsilon d};\epsilon\right) = 0$
at any order of $\epsilon$.

\section{Linear stability of gap solitons} \lsect{stability}

Stability of solitons with respect to perturbations is an important
problem for applications. Stable states act as attractors, and 
their excitation is
weakly sensitive to noise or perturbations. On  the other hand,
unstable states tend to undergo dynamical transformations due to a
rapid growth of initial perturbations, and this behavior may be
useful, for example, for switching
applications~\cite{Morandotti:1999-2726:PRL}.

We study the stability of gap solitons $\Phi_s(x)$ by considering
the evolution of perturbed solution in the following form,
\begin{eqnarray} \nonumber
      \psi(x,t) = e^{-i \mu_s t} \left[ \Phi_s(x)
      + \left( u(x) + i w(x) \right) e^{\lambda t} \right. \\
      \label{linearization} + \left. \left(
      \overline{u(x)} + i \overline{w(x)} \right) e^{\bar{\lambda} t}
      \right].
\end{eqnarray}
We substitute Eq.~(\ref{linearization}) into the NLS equation
(\ref{NLS}) and perform its linearization with respect to the
functions $(u,w)$ describing small-amplitude perturbations. Then,
we obtain coupled linear eigenmode equations where $(u,w)$ is an
eigenvector and $\lambda$ is an eigenvalue,
\begin{equation} \label{eigenvalue}
   {\cal L}_1 u = - \lambda w, \qquad {\cal L}_0 w
   = \lambda u.
\end{equation}
Here ${\cal L}_0$ and ${\cal L}_1$ are Schr\"{o}dinger operators
with periodic and decaying potentials,
\begin{eqnarray} \label{L0}
   {\cal L}_0 & = & - \frac{d^2}{dx^2}
   + V(x) - \mu_s + \sigma \Phi_s^2(x), \\
   \label{L1} {\cal L}_1 & = & - \frac{d^2}{dx^2} + V(x) - \mu_s + 3
   \sigma \Phi_s^2(x).
\end{eqnarray}

We are interested in eigenvalues $\lambda$, which correspond to
the spatially localized eigenvectors $(u,w)$ in $L^2({\mathbb
R},{\mathbb C}^2)$. If there exists an eigenvalue $\lambda$ with
${\rm Re}(\lambda) > 0$, the gap soliton $\Phi_s(x)$ is spectrally
unstable. On contrary, if all eigenvalues have ${\rm Re}(\lambda)
= 0$, the gap soliton is neutrally stable. Neutral stability can
result in spectral instability due to resonances, embedded
eigenvalues, and bifurcations of isolated eigenvalues with ${\rm
Re}(\lambda) = 0$. We have used an approach based on the Evans
function for numerical calculation of the eigenvalues, the details
are presented in Appendix~\rsect{Evans}.

The stability problem (\ref{eigenvalue}) is written in terms of
two Schr\"{o}dinger operators ${\cal L}_0$ and ${\cal L}_1$ with
periodic $V(x)$ and decaying $\sigma \Phi_s^2(x)$ potentials. At
the band edge $\mu = \mu_n$, where $\Phi_s(x) \equiv 0$, the two
Schr\"{o}dinger operators coincide with the operator ${\cal L}_s$:
\begin{equation}
   {\cal L}_s = - \frac{d^2}{d x^2} + V(x) - \mu_s.
\end{equation}
For $w = i u$ and $\lambda = i \Omega$, the spectral bands of the
stability problem (\ref{eigenvalue}) occur at $\Omega + \mu_s \in
\Sigma_{\rm band}$, i.e. at $\Omega \in [\mu_0 -
\mu_s,\mu_1 - \mu_s] \cup [\mu_3 - \mu_s,\mu_2 - \mu_s] \cup ...$.
For the gap soliton bifurcating from the upper band edge $\mu_s =
\mu_n$, the parameter $\mu_s$ satisfies the inequality: $\mu_n <
\mu_s < \mu_{n+2}$, while for the gap soliton bifurcating from the
lower band edge $\mu_s = \mu_n$, the parameter $\mu_s$ satisfies
the inequality: $\mu_{n-2} < \mu_s < \mu_{n}$.

We demonstrate below that an important value which defines many
stability properties is the energy of the spectral band, which is
defined by
\begin{equation} \label{secondvariation}
   h = \langle u, {\cal L}_1 u \rangle_d +
   \langle w, {\cal L}_0 w \rangle_d,
\end{equation}
where the inner product $\langle \cdot,\cdot\rangle_d$ is defined
for periodic Bloch functions on the period $x \in [0,d]$:
\begin{equation} \label{inner-product-period}
   \langle f,g\rangle_d = \int_{0}^d
   \overline{f(x)} g(x) dx.
\end{equation}
It is clear that $h_m = 2 (\mu_m - \mu_n) \langle \psi_m, \psi_m
\rangle_d$ at $\epsilon = 0$, where $h_m$ refers to the $m$-th
band edge in the spectrum of ${\cal L}_s$ for $\mu_s = \mu_n$. All
spectral bands of ${\cal L}_s$, which are lower with respect to
$\mu_s = \mu_n$, become bands of negative energy for the gap
soliton $\Phi_s(x)$, while all spectral bands of ${\cal L}_s$,
which are upper with respect to $\mu_s = \mu_n$, become bands of
positive energy for the gap soliton $\Phi_s(x)$.

The spectrum $\lambda$ of the stability problem (\ref{eigenvalue})
is double because of the inversion symmetry: $w = -iu$ and
$\lambda = -i\Omega$. As a result, the bands of positive and
negative energies of the operators ${\cal L}_s$ and $(-{\cal
L}_s)$ may overlap in the coupled spectrum (\ref{eigenvalue}) for
the same values of $\lambda$.

The spectrum $\lambda$ of the problem (\ref{eigenvalue})
transforms when $\epsilon \neq 0$. A simple and stable
transformation is a shift of spectral bands of ${\cal L}_s$ and
$(-{\cal L}_s)$ along the imaginary axis of $\lambda$ to the
distance $|\mu_s - \mu_n|$. As a result, the origin $\lambda = 0$
becomes isolated from the spectral bands of ${\cal L}_s$ and
$(-{\cal L}_s)$ for any $\epsilon \neq 0$.

Other transformations of the spectrum $\lambda$ are possible
and may result in instabilities of gap solitons. These transformations
are considered in Sections~\rsect{splitting} and ~\rsect{internal_mode}.

\section{Symmetry-breaking instability of gap solitons}
\lsect{splitting}

In Sec.~\rsect{branches}, we have identified two families of
on-site and off-site gap solitons, which have different
positions with respect to the underlying potential. In this
section, we demonstrate that one of these soliton families is
unstable with respect to symmetry breaking. They tend to move
across the potential and eventually transform into their stable
counterparts which have a different position. These
results generalize the previously found instability of off-site
discrete solitons associated with the first
band~\cite{Morandotti:1999-2726:PRL}.

More specifically, we show that the symmetry-breaking 
instability of gap solitons is 
defined by the sign of $M_s'(x_0)$. If $M_s'(x_0) > 0$, then a
pair of purely imaginary eigenvalues $\lambda$ in the stability
problem (\ref{eigenvalue}) bifurcates from $\lambda = 0$, 
and these internal modes describe oscillations of the perturbed
soliton around the stable position $x=x_0$. On the
other hand, if $M_s'(x_0) < 0$, then a pair of real 
eigenvalues $\lambda$ bifurcates in the problem (\ref{eigenvalue}) 
and these exponentially growing instability modes
characterize soliton motion away from the unstable
location $x=x_0$. We note that these results are valid in the vicinity
of gap edges, where the eigenvalues $\lambda$ are exponentially
small in terms of the perturbation parameter $\epsilon$.

Due to the symmetry of the NLS equation (\ref{NLS}), we have
a non-empty kernel of the operator ${\cal L}_0$ for all $\epsilon$
along the family of the gap soliton $\Phi_s(x)$:
\begin{equation}
\label{zero-L0} {\cal L}_0 \Phi_{\epsilon}(x;X) = 0.
\end{equation}
On the other hand, the gap soliton $\Phi_s(x)$ in the asymptotic
representation (\ref{soliton-form}) is parameterized by $x_0$ in
the formal power series (\ref{perturbation3}) in $\epsilon$. As a
result, the kernel of the operator ${\cal L}_1$ is non-empty at
all power orders of $\epsilon^n$:
\begin{equation} \label{null-L1}
   {\cal L}_1 U_{\epsilon} = 0(\epsilon^n), \qquad U_{\epsilon}
   = \frac{\partial \Phi_{\epsilon}(x;X)}{\partial X}.
\end{equation}
The zero eigenvalue of ${\cal L}_1$ is destroyed
beyond the powers of $\epsilon$, since the gap soliton $\Phi_s(x)$
is not parameterized by $x_0$, values of which are fixed by roots
of the Melnikov function (\ref{constraint}). We show in Appendix~\rsect{appA}
that the zero eigenvalue of ${\cal L}_1$, associated with the
eigenfunction $U_{\epsilon}(x)$, shifts according to the quadratic form:
\begin{equation} \label{final-quadratic-form}
   \left( U_{\epsilon}, {\cal L}_1 U_{\epsilon} \right) =
   \frac{1}{2 \epsilon^4} M_s'(x_0),
\end{equation}
where the quadratic form is defined for decaying functions on the
whole line of $x$:
\begin{equation} \label{innerproduct}
     \left(f, g \right) = \int_{-\infty} \overline{f(x)} g(x) dx.
\end{equation}

According to the standard perturbation
theory~\cite{Horn:1985:MatrixAnalysis}, the quadratic form in
(\ref{final-quadratic-form}) determines the shift of the zero
eigenvalue of ${\cal L}_1$, associated with the eigenfunction
$U_{\epsilon}(x)$. When $M_s'(x_0) > 0$, the zero eigenvalue of
${\cal L}_1$ becomes positive, while when $M_s'(x_0) < 0$, the
zero eigenvalue of ${\cal L}_1$ becomes negative. We show that a
small negative eigenvalue of ${\cal L}_1$ results in a small real
positive eigenvalue $\lambda$ of the stability problem
(\ref{eigenvalue}), while a small positive eigenvalue ${\cal L}_1$
results in a pair of small imaginary eigenvalues $\lambda$.

A small eigenvalue $\lambda = \lambda_{\epsilon}$, corresponding
to the eigenfunction $u_{\epsilon}(x)$, can be found from the
problem:
\begin{equation} \label{eigenvalue0}
   {\cal L}_1 u_{\epsilon} = -
   \lambda_{\epsilon}^2 {\cal L}_0^{-1} u_{\epsilon},
\end{equation}
or equivalently, from the Rayleigh quotient:
\begin{equation} \label{eigenvalue-approximation}
   \lambda_{\epsilon}^2 = -
   \frac{\left( u_{\epsilon}, {\cal L}_1 u_{\epsilon}
   \right)}{\left( u_{\epsilon}, {\cal L}_0^{-1} u_{\epsilon}
   \right)}.
\end{equation}
The quadratic form $\left( u_{\epsilon}, {\cal L}_0^{-1}
u_{\epsilon}\right)$ exists if $(\Phi_{\epsilon},u_{\epsilon})
= 0$, as follows from (\ref{zero-L0}). Since
$(\Phi_{\epsilon},U_{\epsilon}) = {\rm O}(\epsilon^n)$ 
and ${\cal L}_1 U_{\epsilon} = {\rm O}(\epsilon^n)$ 
at all power orders of $\epsilon^n$, we conclude that
\begin{equation}
   u_{\epsilon}(x) = U_{\epsilon}(x) + E_{\epsilon},
\end{equation}
where $E_{\epsilon}$ is exponentially small in terms of
$\epsilon$. We shall prove that
\begin{eqnarray} \label{positive-quadratic-form}
   \left( u_{\epsilon}, {\cal L}_0^{-1} u_{\epsilon}
   \right) = \frac{1}{4 \epsilon^2} \left( \Phi_{\epsilon},
   \Phi_{\epsilon} \right) + {\rm
   O}\left(\frac{1}{\epsilon}\right),
\end{eqnarray}
such that $\left( u_{\epsilon}, {\cal L}_0^{-1} u_{\epsilon}
\right) > 0$ at the leading order. It follows from the nonlinear
problem (\ref{ODE}) that
\begin{equation}
   {\cal L}_0 X \Phi_{\epsilon}(x;X) = - 2 \epsilon \frac{\partial
   \Phi_{\epsilon}(x;X)}{\partial x} - 2 \epsilon^2 \frac{\partial
   \Phi_{\epsilon}(x;X)}{\partial X}.
\end{equation}
As a result, we have
\begin{eqnarray} \nonumber
    -\frac{1}{2 \epsilon^2} \left( \frac{\partial
   \Phi_{\epsilon}}{\partial X}, X \Phi_{\epsilon} \right) \\
   \label{lengthy-computations}
   = \left( \frac{\partial
   \Phi_{\epsilon}}{\partial X}, {\cal L}_0^{-1}
   \frac{\partial \Phi_{\epsilon}}{\partial X} \right) +
   \frac{1}{\epsilon} \left( \frac{\partial
   \Phi_{\epsilon}}{\partial X}, {\cal L}_0^{-1}
   \frac{\partial \Phi_{\epsilon}}{\partial x} \right).
\end{eqnarray}
Solution of the inhomogeneous problem,
\begin{equation} \label{equations-v-L0}
   {\cal L}_0 V_{\epsilon} = \frac{\partial
   \Phi_{\epsilon}}{\partial x}, 
\end{equation}
exists at all power orders of $\epsilon^n$, since the
right-hand-side of (\ref{equations-v-L0}) is orthogonal to
$\Phi_{\epsilon}$ at all power orders of $\epsilon^n$. Therefore,
the quadratic form $(U_{\epsilon},V_{\epsilon})$ has a 
regular power series in $\epsilon$, starting with 
the zero-order term. Since
\begin{eqnarray}
\nonumber
 -\frac{1}{2 \epsilon^2} \left( \frac{\partial \Phi_{\epsilon}}{\partial X}, X
\Phi_{\epsilon} \right) & = & \frac{1}{4 \epsilon^2} \left(
\Phi_{\epsilon},\Phi_{\epsilon} \right) \\
& - & \frac{1}{4 \epsilon^2}
\int_{-\infty}^{\infty} \frac{\partial}{\partial X} 
\left( X \Phi^2_{\epsilon} \right) dx,
\end{eqnarray}
and the second term is exponentially small in $\epsilon$, we have
(\ref{positive-quadratic-form}) at the leading order, such that
the Rayleigh quotient (\ref{eigenvalue-approximation}) is given in
the leading order by:
\begin{equation} \label{perturbation-theory}
   \lambda^2_{\epsilon} \approx - \frac{2
   M_s'(x_0)}{\epsilon^2 \left( \Phi_{\epsilon}, \Phi_{\epsilon}
   \right)}.
\end{equation}
It follows from (\ref{perturbation-theory}) that
a negative eigenvalue of ${\cal L}_1$ for
$M_s'(x_0) < 0$ results in a small positive eigenvalue
$\lambda_{\epsilon}$ in the stability problem (\ref{eigenvalue}).

The exponentially small correction of the function $M_s(x_0)$ is
given by the expansion (\ref{roots-bif}), where ${\rm
arg}\left(w_{n,1}^{(0)}\right) = - \frac{\pi}{2}$ for the
square-sine potential (\ref{sine}). Therefore, $M_s'(x_0) > 0$ for
$x_0 = 0$ and $M_s'(x_0) < 0$ for $x_0 = \frac{d}{2}$. In the
former case, the gap soliton $\Phi_s(x)$ is located at the minimum
point of $V(x)$ and it has a pair of small imaginary eigenvalues
$\lambda$. In the latter case, the gap soliton $\Phi_s(x)$ is
located at the maximum point of $V(x)$ and it is unstable with a
small real positive eigenvalue $\lambda$. Fig.~\rpict{instabReal}
shows unstable eigenvalues, splitting from zero eigenvalues, for
the branches of gap solitons with $x_0 = \frac{d}{2}$. We note
that stability of on-site and off-site solitons can be
interchanged in more complex potentials, such as binary
superlattices~\cite{Sukhorukov:2003-2345:OL}. Additionally,
stability can change deep inside the gap, where the asymptotic
analysis is not applicable~\cite{Gorbach:2004-77:EPD}.

\pict{fig04.eps}{instabReal}{ Eigenvalues corresponding to
symmetry-breaking instabilities of gap solitons centered at $x_0=d/2$
(shown with dashed lines in Fig.~\rpict{solitonSF}) for $\sigma =
-1$. }

Asymptotic results for NLS solitons in the lowest semi-infinite
band gap in the focusing case ($\sigma = -1$) were obtained
recently by Kapitula~\cite{Kapitula:2001-186:PD} in the limit
$V(x) \to 0$. Branches of NLS solitons $\Phi_s(x) = \Phi_{\rm
NLS}(x) = \sqrt{2\mu_s} \; {\rm sech}(\sqrt{2\mu_s}(x-x_0))$ in
the small periodic potential function $V(x)$ are defined by zeros
of the function $M_s(x_0)$, given by (\ref{constraint}) with
$\Phi_s = \Phi_{\rm NLS}(x)$. Stability of branches of NLS
solitons is defined by the derivative $M_s'(x_0)$, such that the
NLS solitons bifurcating from the minimum points of $V(x)$ are
stable, while the NLS solitons bifurcating from the maximum points
of $V(x)$ are unstable. We note that the opposite conclusion is
drawn in Ref.~\cite{Kapitula:2001-186:PD}, due to an elementary
sign error.

\section{Internal modes and oscillatory instabilities of gap solitons}
\lsect{internal_mode}

Apart from symmetry-breaking instabilities analyzed in the
previous section, we demonstrate that gap solitons can exhibit a
different type, so-called oscillatory instabilities. 
Such instabilities can occur due to a resonance between the modes
corresponding to the edges of the gap in which soliton is
localized, as was demonstrated in the case of a narrow
gap when the coupled-mode equations are
applicable~\cite{Barashenkov:1998-5117:PRL}. However, the
coupled-mode theory described only an isolated band-gap, whereas
it was found that oscillatory instability can occur due to
resonance between different gaps~\cite{Sukhorukov:2001-83901:PRL,
Sukhorukov:2003-2345:OL}. Such resonances can result in 
energy redistribution between the gaps and a formation of
breathing structures, as was recently demonstrated
experimentally~\cite{Neshev:nlin.PS/0311059:ARXIV}. In this
section, we present a systematic analysis of such instabilities,
and show that they appear when a side-band associated with the
inter-gap resonances falls outside a band-gap.

Oscillatory modes and instabilities are characterized by
eigenvalues $\lambda$ with non-zero imaginary part of the
stability problem (\ref{eigenvalue}) for $\epsilon \neq 0$. First,
we show that new imaginary eigenvalues  $\lambda$ with decaying
eigenvectors $(u,w)$ bifurcate from the band edges $\lambda = i
(\mu_m - \mu_n)$ of the same polarity as the band edge $\mu_s =
\mu_n$. Bifurcations of internal modes occur generally at the
order of ${\rm O}(\epsilon^2)$, if $\mu_m \neq \mu_n$. These
eigenvalues are referred to as the internal modes of gap
solitons~\cite{Pelinovsky:1998-121:PD, Kivshar:1998-5032:PRL}, and
in our case such modes appear due to a resonance between the gap
edges $m$ and $n$. Such resonances are possible because a soliton
induces an effective waveguide, which can support localized modes
in other gaps~\cite{Cohen:2003-113901:PRL, Sukhorukov:2003-113902:PRL}. In
Fig.~\rpict{modesL0}, we show three modes of operator 
${\cal L}_0$ supported in the semi-infinite gap near the edge 
$\mu=\mu_0$ by a gap soliton existing in the 
gap near the edge $\mu=\mu_2$ in the case
of a self-focusing nonlinearity ($\sigma=-1$).

\pict{fig05.eps}{modesL0}{Linear guided modes of operator 
${\cal L}_0$ in the semi-infinite band-gap for a gap
soliton shown in Fig.~\rpict{solitonSF}(d). Left:~eigenvalues
marked with dots (second and third ones are indistinguishable
within the picture scale); Right:~corresponding mode profiles. }

Second, we show that resonance between internal modes of the
operator ${\cal L}_s$ and the bands of the inverted spectrum of
$(-{\cal L}_s)$ occurs if the bifurcating internal mode of ${\cal
L}_s$ becomes embedded into the spectral band of $(-{\cal L}_s)$.
When it happens, embedded internal modes bifurcate generally to
complex eigenvalues $\lambda$, leading to oscillatory
instabilities of the gap soliton $\Phi_s(x)$. Resonant
bifurcations of complex eigenvalues $\lambda$ occur generally at
order of ${\rm O}(\epsilon^4)$.

Third, we show that the internal mode of ${\cal L}_s$ may occur
near the band edge of the inverted spectrum of $(-{\cal L}_s)$. In
this case, bifurcations of isolated, embedded, and complex
eigenvalues are all possible at the order of ${\rm
O}(\epsilon^2)$, depending on the configuration of the spectral
bands of ${\cal L}_s$ and $(-{\cal L}_s)$.

Finally, we show that at most one internal mode can bifurcate from the band
edge, which is closest to the zero eigenvalue. This bifurcation
occurs generally at the order of ${\rm O}(\epsilon^4)$.

We emphasize that bifurcations of new eigenvalues may not
generally occur in higher orders of $\epsilon$, since the band
edges of the spectrum of ${\cal L}_1$ and ${\cal L}_0$ with
$\Phi_s(x) \neq 0$ do not support resonances in a generic case.
Bifurcations of new eigenvalues may occur far from the limit
$\epsilon = 0$, when the spectral bands of the linearized operator
get additional resonances at the band edges or in the interior
points. Bifurcations of the existing eigenvalues may also occur
far from the limit $\epsilon = 0$, if the existing eigenvalues
coalesce with each other or with the spectral bands.

\subsection{Non-resonant bifurcations of internal modes}

Let $n$ be the index of the band edge $\mu_s = \mu_n$ where the
gap soliton $\Phi_s(x)$ bifurcates from. We consider a different
band edge of the stability problem (\ref{eigenvalue}) with
$\lambda = i(\mu_m - \mu_n)$, such that $m \neq n$. We assume
that the $m$-th band edge of the spectrum of ${\cal L}_s$ is
located in a band gap of the inverted spectrum of $(-{\cal L}_s)$,
such that $2 \mu_n - \mu_m \notin \Sigma_{\rm band}$.
Using the same perturbation series expansions (\ref{perturbation2}) and
(\ref{perturbation3}), we expand solutions of the stability
problem (\ref{eigenvalue}) in the perturbation series:
\begin{eqnarray} \nonumber
     u = B_m(X) \psi_m(x) + \epsilon B_m'(X)
    \psi_m^{(1)}(x) \\ \label{perturbation4a}
    + \epsilon^2 u_m^{(2)}(x,X) + {\rm O}(\epsilon^3), \\
    \nonumber w = i \left[ B_m(X) \psi_m(x) +
    \epsilon B_m'(X) \psi_m^{(1)}(x) \right. \\ \label{perturbation4b}
    + \left. \epsilon^2 w_m^{(2)}(x,X) + {\rm O}(\epsilon^3) \right],
\end{eqnarray}
and
\begin{equation} \label{perturbation4}
   \lambda = i \left[ \mu_m - \mu_s +
   \epsilon^2 \Omega_m + {\rm O}(\epsilon^4) \right],
\end{equation}
where the second-order correction terms $(u_m^{(2)},w_m^{(2)})$
solve the non-homogeneous system:
\begin{eqnarray} \nonumber
   {\cal L}_s u_m^{(2)} + (\mu_s - \mu_m)
   w_m^{(2)} = B_m'' \left( \psi_m + 2 \left( \psi_m^{(1)}
   \right)' \right) \\
    \label{nonhomgen1} + \Omega_m B_m \psi_m - 3 \sigma A_n^2 B_m
   \psi_n^2 \psi_m, \\
   \nonumber
   {\cal L}_s w_m^{(2)} + (\mu_s - \mu_m)
   u_m^{(2)} = B_m'' \left( \psi_m + 2 \left( \psi_m^{(1)}
   \right)' \right) \\
    \label{nonhomgen2} + \Omega_m B_m \psi_m - \sigma A_n^2 B_m \psi_n^2
   \psi_m.
\end{eqnarray}
Under the constraint that $(2 \mu_s - \mu_m) \notin \Sigma_{\rm
band}$, the second-order corrections $(u_m^{(2)},w_m^{(2)})$ are
periodic or anti-periodic functions of $x$, when a single Fredholm
condition is satisfied. The Fredholm
condition takes the form of the eigenvalue problem for $\Omega_m$:
\begin{equation} \label{waveguide}
   \mu_m^{(2)} B_m'' + 2 \chi_{nm}^{(2)} A_n^2(X)
   B_m - \Omega_m B_m = 0,
\end{equation}
where
\begin{equation} \label{chinm2}
   \chi_{nm}^{(2)} = \sigma \frac{\int_0^d \psi_n^2
   \psi_m^2 dx}{\int_0^d \psi_m^2 dx}.
\end{equation}
We note that Eq.~(\ref{waveguide}) describes the linear modes
supported by a soliton-induced waveguide in other gaps. The linear
problem (\ref{waveguide}) is a Schr\"{o}dinger equation with the
solvable potential (\ref{NLSsoliton}). There is
at least one isolated eigenvalue if ${\rm sign}(\mu_m^{(2)}) = {\rm
sign}(\chi_{nm}^{(2)}) = {\rm sign}(\Omega_m)$. In this case, the
lowest eigenvalue and eigenfunction of the problem
(\ref{waveguide}) can be found explicitly as
\begin{equation}
   \Omega_m = \kappa_n^2 \mu_m^{(2)} s_m^2, \qquad
   B_m = {\rm sech}^{s_m}(\kappa_n X),
\end{equation}
where $s_m$ solves the quadratic equation,
\begin{equation}
   s_m (s_m + 1) = \frac{4 \mu_n^{(2)} \chi_{nm}^{(2)}}{\mu_m^{(2)}
   \chi_n^{(2)}}.
\end{equation}
Isolated eigenvalues $\Omega_m$ of the problem (\ref{waveguide}),
when exist, correspond to internal modes $\lambda$ in the
perturbation series (\ref{perturbation4}), bifurcating in the band
gaps of the operators ${\cal L}_s$ and $(-{\cal L}_s)$ from the
band edge $\lambda = i(\mu_m - \mu_n)$. When ${\rm
sign}(\mu_m^{(2)}) = -{\rm sign}(\chi_{nm}^{(2)})$, the linear
problem (\ref{waveguide}) does not have any isolated eigenvalues.
Since ${\rm sign}(\chi_{nm}^{(2)}) = {\rm sign}(\chi_n^{(2)}) =
{\rm sign}(\sigma)$, we notice that all band edges $\mu = \mu_n$
that support bifurcations of gap solitons in the nonlinear problem
(\ref{ODE}), support also bifurcations of internal modes $\lambda$
in the spectrum of a selected $n$-th gap soliton. In the focusing
case, $\sigma = -1$, all lower band edges generate internal modes
$\lambda$ downwards the corresponding band gaps, i.e.
$\mu_{m}^{(2)} < 0$ and $\Omega_m < 0$. In the defocusing case,
$\sigma = +1$, all upper band edges generate internal modes
$\lambda$ upwards the corresponding band edges, i.e.
$\mu_{m}^{(2)} > 0$ and $\Omega_m > 0$.

It is surprising that more than one internal mode $\lambda$ could
be generated near the band edge $\lambda = i(\mu_m - \mu_n)$. 
In the case of no periodic potential $V(x) = 0$,
perturbations of NLS solitons generate at most one internal
mode~\cite{Kivshar:1998-5032:PRL, Kapitula:2001-533:NLN}. On the
other hand, perturbations of gap solitons in the coupled-mode
equations (derived for small $V(x)$ in the narrow band gap of the
first resonance) may generate several internal modes and complex
eigenvalues~\cite{Barashenkov:1998-5117:PRL,
Kapitula:2002-1117:SJMA}. In the case of finite potential $V(x)$,
the number of internal modes depends on the depth of the squared
sech potential in the eigenvalue problem (\ref{waveguide}), which
is determined by parameters of the band curvatures $\mu_n^{(2)}$
and $\mu_m^{(2)}$ and by the nonlinearity coefficients
$\chi_n^{(2)}$ and $\chi_{nm}^{(2)}$.

\subsection{Resonant bifurcations of complex eigenvalues}

According to the general expression~(\ref{linearization}),
eigenvalues $\lambda$ with non-zero imaginary part describe
soliton oscillations, which are associated with the appearance of
two side-band spatial frequencies $\mu+{\rm Im}(\lambda)$ and
$\mu-{\rm Im}(\lambda)$. Gap solitons are spectrally stable for
small values of $\epsilon \neq 0$ with respect to a particular
resonant oscillation if both of the side-bands fall inside the
gaps of the linear spectrum, whereas an oscillatory instability
may arise when one side-band appears inside a linear transmission
band~\cite{Sukhorukov:2001-83901:PRL, Sukhorukov:2003-2345:OL}.
This general behavior is illustrated in
Fig.~\rpict{instabOscilSF2}, where the real part of the eigenvalue
is non-zero indicating a presence of the oscillatory instability
when the lower side-band is inside the transmission band 
of the inverted spectrum. However, the
instability is suppressed when the side-band moves 
inside the band gap.
The instability shown in Fig.~\rpict{instabOscilSF2} appears due
to a resonant coupling between a gap soliton marked "d" in
Fig.~\rpict{solitonSF}, and its own fundamental guided mode in the
first gap shown in Fig.~\rpict{modesL0}. The characteristic
profiles of instability modes are presented in
Fig.~\rpict{instabOscilSF2prof}. The top row shows the
perturbation $u+i w$, which corresponds to higher spatial
frequency $\mu+{\rm Im}(\lambda)$ according to
Eq.~(\ref{linearization}), and we indeed see that this mode
closely matches the guided mode profile [cf. Fig.~\rpict{modesL0}]
in agreement with the asymptotic expressions~(\ref{perturbation4a}) and~(\ref{perturbation4b}). On the other hand, the bottom
row of Fig.~\rpict{instabOscilSF2prof} shows the low-frequency
component, which describes the radiation waves emitted by the
soliton when $\mu-{\rm Im}(\lambda)$ is inside the transmission 
band. We
note however that the rate of such radiation may be very small,
allowing for the existence of long-lived oscillating, or breathing
states, such as shown in Fig.~\rpict{instabOscilSF2_bpm}. Similar
effects may occur due to a resonance with higher-order guided
modes, as shown in
Figs.~\rpict{instabOscilSF2b}-\rpict{instabOscilSF2b_bpm}. One
important difference is that the associated breathing states can
have different symmetries for various excited modes, cf.
Figs.~\rpict{instabOscilSF2_bpm} and~\rpict{instabOscilSF2b_bpm}.

\pict{fig06.eps}{instabOscilSF2}{ Eigenvalues corresponding
to a resonance of a gap soliton (marked "d" in
Fig.~\rpict{solitonSF}) with its fundamental guided mode in the
semi-infinite band-gap. }

\pict{fig07.eps}{instabOscilSF2prof}{ Profiles of linear
modes corresponding to a resonance in Fig.~\rpict{instabOscilSF2}.
}

\pict{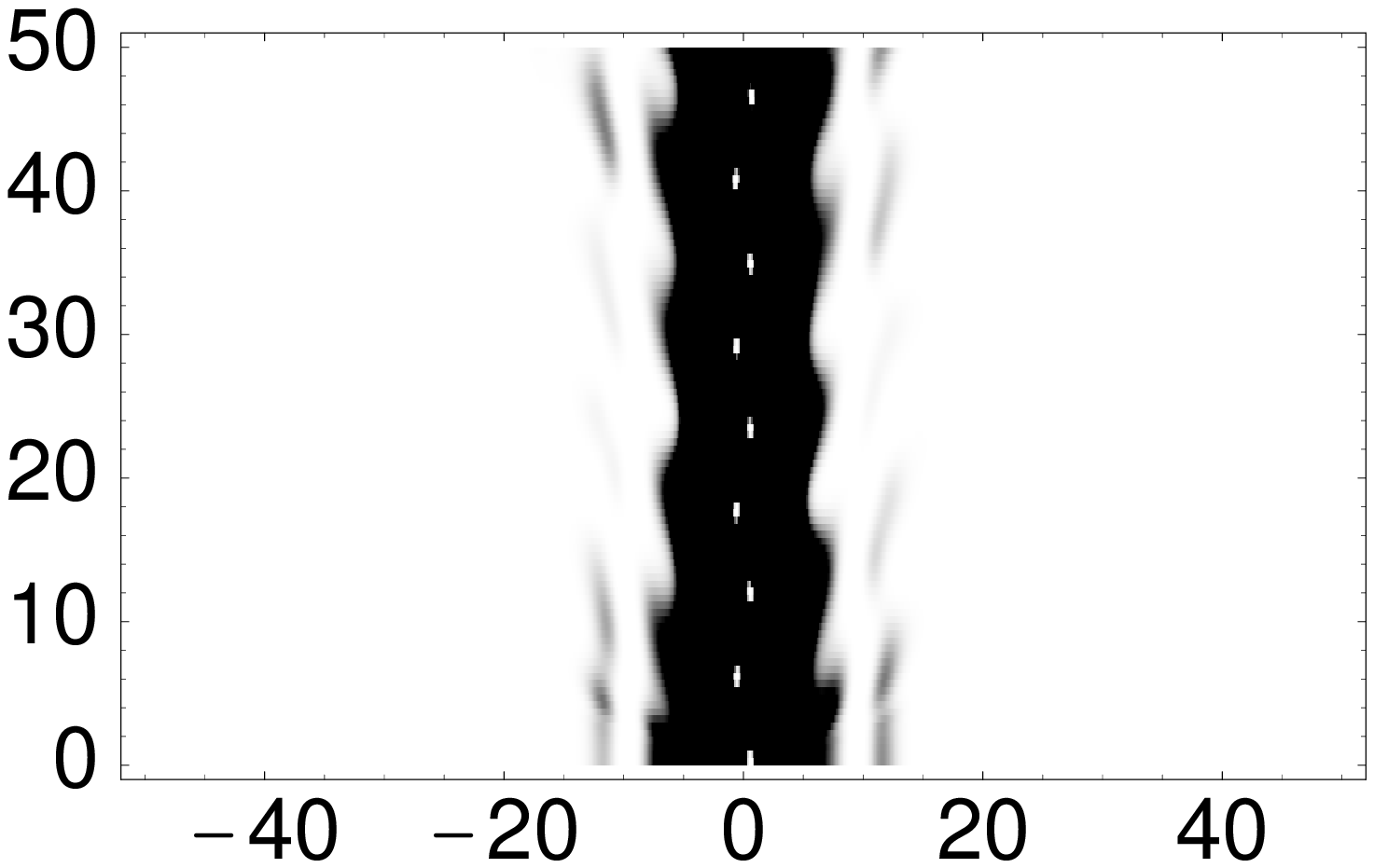}{instabOscilSF2_bpm}{ Evolution of a
soliton perturbed by a linear mode corresponding to
Fig.~\rpict{instabOscilSF2prof}. }

\pict{fig09.eps}{instabOscilSF2b}{ Eigenvalues corresponding
to a resonance of a gap soliton (marked "d" in
Fig.~\rpict{solitonSF}) with its higher-order guided mode in the
semi-infinite band-gap. }

\pict{fig10.eps}{instabOscilSF2bprof}{ Profiles of
linear modes corresponding to a resonance in
Fig.~\rpict{instabOscilSF2b}. }

\pict{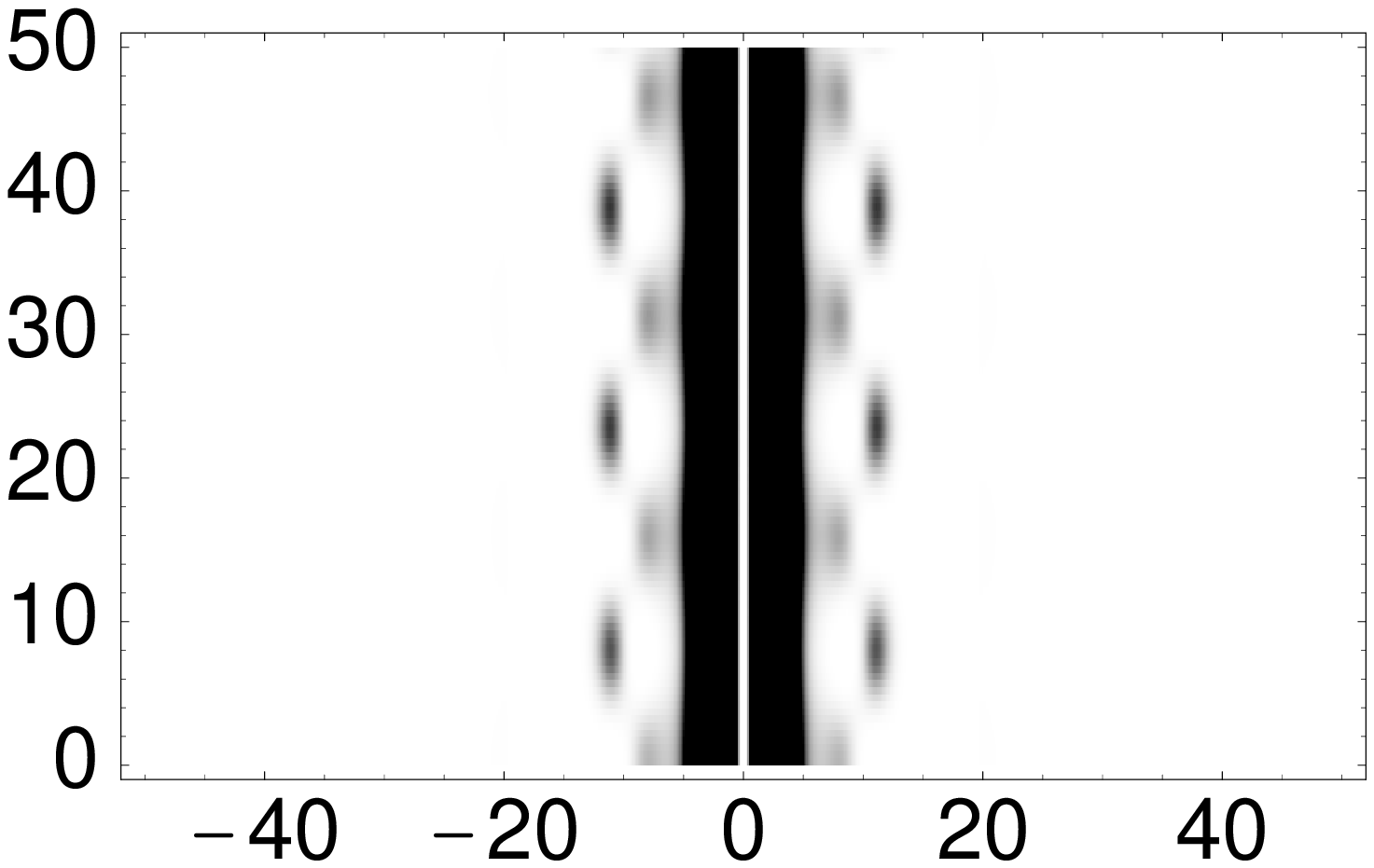}{instabOscilSF2b_bpm}{ Evolution of a
soliton perturbed by a linear mode corresponding to
Fig.~\rpict{instabOscilSF2bprof}. }

In mathematical terms, stability requires that all internal modes
detaching from the band edges $\lambda = i (\mu_m - \mu_n)$ reside
inside the band gaps of the inverted operator $(-{\cal L}_s)$, and
the zero eigenvalue of ${\cal L}_1$ shifts to small imaginary
eigenvalues $\lambda$. When an internal mode is embedded into a
spectral band of the inverted operator $(-{\cal L}_s)$,
oscillatory instability of the gap soliton $\Phi_s(x)$ may arise
for small values of $\epsilon \neq 0$. Embedded imaginary
eigenvalues $\lambda$ are known to be structurally unstable with
respect to small perturbations and, provided that their energy is
opposite with respect to the energy density of the spectral band,
they bifurcate into complex eigenvalues
$\lambda$~\cite{Tsai:2002-1629:IMRN}. By construction, resonance
of internal modes of ${\cal L}_s$ with spectral bands of $(-{\cal
L}_s)$ is only possible if the internal mode, detaching from the
band edge $\lambda = i (\mu_m - \mu_s)$, has the opposite energy
signature with respect to the energy signature of the inverted
spectral band $\mu_r = 2 \mu_n - \mu_m \in \Sigma_{\rm band}$,
such that $\lambda = i (\mu_m - \mu_n) = i( \mu_n - \mu_r)$.
Therefore, all embedded imaginary eigenvalues in the linearized
stability problem (\ref{eigenvalue}) are expected to bifurcate to
complex eigenvalues $\lambda$ in a generic case.

We prove in Appendix~\rsect{appB} that, provided that $\mu_r = 2
\mu_n - \mu_m \in \Sigma_{\rm band}$, we have
\begin{equation} \label{final-Fermi-rule}
   {\rm Re}(\lambda) = \epsilon^4 \Gamma_m + {\rm O}(\epsilon^5), \qquad
   \Gamma_m \ge 0,
\end{equation}
where $\lambda$ is the eigenvalue of the bifurcating internal
mode, given by (\ref{perturbation4}). In a generic case, when
$\Gamma_m \neq 0$, the embedded imaginary eigenvalue $\lambda$
bifurcates to the unstable domain ${\rm Re}(\lambda) > 0$ and
leads to oscillatory instabilities of the gap soliton $\Phi_s(x)$.

\subsection{Marginal bifurcations of internal modes and complex eigenvalues}

A marginal case between non-resonant and resonant bifurcations
occurs when internal modes detaching from the band edge $\lambda =
i(\mu_m - \mu_n)$ are located in the neighborhood of the band edge
$\lambda = i(\mu_n - \mu_k)$ of the inverted spectrum. We assume
here that $\mu_m$, $\mu_k$, and $\mu_s$ satisfy the resonance
condition within the mismatch of order ${\rm O}(\epsilon^2)$:
\begin{equation} \label{resonance}
   \mu_m + \mu_k - 2 \mu_s = \epsilon^2 \nu_{mk}.
\end{equation}
In this marginal case, we expand the eigenvalue $\lambda$ and the
eigenfunction $(u,w)$ of the linearized stability problem
(\ref{eigenvalue}) in the modified perturbation series,
\begin{eqnarray}
    \nonumber u = u_{mk}^{(0)}(x;X) + \epsilon
   u_{mk}^{(1)}(x;X) \\
    \label{perturbation7}
    + \epsilon^2 u_{mk}^{(2)}(x;X) + {\rm
   O}(\epsilon^3), \\ \nonumber
   w = i \left[ w_{mk}^{(0)}(x;X) + \epsilon
   w_{mk}^{(1)}(x;X) \right. \\  \label{perturbation8}
   + \left. \epsilon^2 w_{mk}^{(2)}(x;X) + {\rm
   O}(\epsilon^3) \right],
\end{eqnarray}
and
\begin{equation} \label{perturbation67}
   \lambda = i \left[ \mu_m - \mu_s +
   \epsilon^2 \Omega_{mk} + {\rm O}(\epsilon^4) \right],
\end{equation}
where
\begin{eqnarray*}
   u_{mk}^{(0)} & = & B_m(X) \psi_m(x) + C_k(X) \psi_k(x), \\
   u_{mk}^{(1)} & = & B_m'(X) \psi_m^{(1)}(x) + C_k'(X)
   \psi_k^{(1)}(x), \\
   w_{mk}^{(0)} & = & B_m(X) \psi_m(x) - C_k(X) \psi_k(x), \\
   w_{mk}^{(1)} & = & B_m'(X) \psi_m^{(1)}(x) - C_k'(X) \psi_k^{(1)}(x).
\end{eqnarray*}
The second-order correction terms
$(u_{mk}^{(2)},w_{mk}^{(2)})$ solve the system:
\begin{eqnarray}  \nonumber
   {\cal L}_s u_{mk}^{(2)} + (\mu_s - \mu_m) w_{mk}^{(2)} & = & F^{(2)}, \\
   \label{nonhomgen12}
   {\cal L}_s w_{mk}^{(2)} + (\mu_s - \mu_m) u_{mk}^{(2)} & = & G^{(2)},
\end{eqnarray}
where
\begin{eqnarray*}
    F^{(2)} & = & B_m'' \left( \psi_m + 2 \left( \psi_m^{(1)} \right)' \right) +
   C_k'' \left( \psi_k + 2 \left( \psi_k^{(1)} \right)' \right) \\
    & + & \Omega_{mk} \left( B_m \psi_m - C_k \psi_k
   \right) - \nu_{mk} C_k \psi_k \\
   & - & 3 \sigma A_n^2 \psi_n^2 \left( B_m \psi_m + C_k \psi_k \right), \\
   G^{(2)} & = & B_m'' \left( \psi_m + 2 \left( \psi_m^{(1)} \right)' \right)
   - C_k'' \left( \psi_k + 2 \left( \psi_k^{(1)} \right)' \right) \\
   & + & \Omega_{mk} \left( B_m \psi_m + C_k \psi_k
   \right) + \nu_{mk} C_k \psi_k \\
   & - & \sigma A_n^2 \psi_n^2 \left( B_m
   \psi_m - C_k \psi_k \right).
\end{eqnarray*}
Because of the resonance condition (\ref{resonance}), the
second-order corrections $(u_{mk}^{(2)},w_{mk}^{(2)})$ are
periodic or anti-periodic functions of $x$ if two Fredholm
conditions are satisfied. The two Fredholm conditions take the
form of a coupled eigenvalue problem for $\Omega_{mk}$:
\begin{eqnarray} \nonumber
   && \mu_m^{(2)} B_m'' + A_n^2(X) \left( 2 \chi_{nm}^{(2)} B_m
   + \chi_{nmk}^{(2)} C_k \right)  = 
    \label{couple1}  \Omega_{mk} B_m, \\
    && \nonumber \mu_k^{(2)} C_k'' + A_n^2(X)
   \left(\chi_{nkm}^{(2)} B_m + 2 \chi_{nk}^{(2)} C_k \right) =  \\
&& \label{couple2} \qquad\qquad\qquad 
- \left( \nu_{mk} + \Omega_{mk} \right) C_k,
\end{eqnarray}
where $\chi_{nm}^{(2)}$ is defined in (\ref{chinm2}), while
$\chi_{nmk}^{(2)}$ and $\chi_{nkm}^{(2)}$ are defined as
\begin{equation}
   \chi_{nmk}^{(2)} = \sigma \frac{\int_0^d \psi_n^2 \psi_m \psi_k
   dx}{\int_0^d \psi_m^2 dx}.
\end{equation}
The coupled eigenvalue problem (\ref{couple1})--(\ref{couple2}) is
not self-adjoint and therefore the eigenvalues $\Omega_{mk}$ could
be complex-valued.

We assume that ${\rm sign}(\mu_m^{(2)}) = {\rm
sign}(\chi_{nm}^{(2)})$ such that the first equation
(\ref{couple1}) with $C_k \equiv 0$ has at least one internal mode
for ${\rm sign}(\Omega_{mk}) = {\rm sign}(\mu_m^{(2)})$. For
convenience, we consider here the defocusing case $\sigma = 1$,
when $\chi_{nm}^{(2)} > 0$ and $\mu_m^{(2)} > 0$. In this case,
the internal mode of the first equation (\ref{couple1}) with $C_k
\equiv 0$ exists for $\Omega_{mk} > 0$, while the spectral band is
located for negative values of $\Omega_{mk}$. There are two
particular cases, depending on whether $\mu_k^{(2)} > 0$ or
$\mu_k^{(2)} < 0$.

In the case $\mu_k^{(2)} < 0$, the second equation (\ref{couple2})
with $B_m \equiv 0$ does not have any internal modes, while the
spectral band is located below the value $\Omega_{mk} \leq
-\nu_{mk}$. When $\nu_{mk} \gg 1$, internal modes in the component
$B_m$ for $\Omega_{mk} > 0$ are not affected by the spectral band
in the component $C_k$, since the following estimate holds for
finite $\Omega_{mk}$ and large $\nu_{mk}$:
\begin{equation} \label{small-C}
   C_k = - \frac{\chi_{nkm}^{(2)}}{\nu_{mk}} A_n^2(X)
   B_m + {\rm O}\left(\frac{1}{\nu_{mk}^2}\right).
\end{equation}
When the value of $\nu_{mk}$ decreases and becomes negative, all
internal modes in the component $B_m$ become embedded into the
spectral band in the component $C_k$. The embedded eigenvalues
$\Omega_{mk}$ bifurcate as complex eigenvalues $\Omega_{mk}$ due
to Fermi golden rule as in~\cite{Tsai:2002-1629:IMRN}.

In the case of $\mu_k^{(2)} > 0$, the second equation
(\ref{couple2}) with $B_m \equiv 0$ has at least one internal mode
for $\nu_{mk} + \Omega_{mk} < 0$, while the spectral band is
located above the value $\Omega_{mk} \geq -\nu_{mk}$. When
$\nu_{mk} \ll -1$, all internal modes in the component $B_m$ for
$\Omega_{mk} > 0$ and those in the component $C_k$ for $\nu_{mk} +
\Omega_{mk} < 0$ are located in the gap between the two spectral
bands. The internal modes in the component $B_m$ are not affected
by the spectral band in the component $C_k$, since $C_k$ is small
according to the expansion (\ref{small-C}). On the other hand, the
internal modes in the component $C_k$ are not affected by the
spectral band in the component $B_m$, since the following estimate
holds for finite $(\Omega_{mk} + \nu_{mk})$ and large
$\Omega_{mk}$:
\begin{equation} \label{small-B}
   B_m = \frac{\chi_{nmk}^{(2)}}{\Omega_{mk}}
   A_n^2(X) C_k + {\rm O}\left(\frac{1}{\Omega_{mk}^2}\right).
\end{equation}
When the value $\nu_{mk}$ increases and becomes positive, the gap
between spectral bands disappear and all internal modes in the
components $B_m$ and $C_k$ coalesce or become embedded into
overlapping spectral bands. In the first case, internal modes
$\Omega_{mk}$ bifurcate as complex eigenvalues $\Omega_m$ due to
the Hamiltonian Hopf bifurcation. In the second case, internal
modes $\Omega_{mk}$ bifurcate as complex eigenvalues $\Omega_{mk}$
due to Fermi golden rule. Again, we have oscillatory instabilities
of the gap soliton $\Phi_s(x)$, emerging from all bifurcating
internal modes of ${\cal L}_s$ in resonance with the spectral
bands of $(-{\cal L}_s)$ or vice verse.

\subsection{Internal modes near $\lambda = 0$}

The coupled eigenvalue problem (\ref{couple1})--(\ref{couple2}) is
derived under the resonance condition (\ref{resonance}) between
two band edges of operators ${\cal L}_s$ and $(-{\cal L}_s)$. The
resonance condition (\ref{resonance}) is always satisfied for
$\mu_m = \mu_k = \mu_n$ and $\nu_{mk} = - 2 \Delta_n$, when the
band edge $\lambda = i(\mu_m - \mu_n) = 0$ of the stability
problem (\ref{eigenvalue}) coincides with the band edge $\mu_s =
\mu_n$ of the gap soliton $\Phi_s(x)$ and $\Delta_n$ is used in
(\ref{perturbation2}). In this case, the coupled
eigenvalue problem (\ref{couple1})--(\ref{couple2}) describes the
transformation of the spectrum of the problem (\ref{eigenvalue})
at $\epsilon \neq 0$, when a narrow spectral gap appears in the
spectrum of the problem (\ref{eigenvalue}) near the origin
$\lambda = 0$.

We showed in Section~\rsect{splitting} that a pair of real
or purely imaginary eigenvalues bifurcate from $\lambda = 0$
due to the broken translational invariance. We will show here
that another pair of internal mode may bifurcate inside the same gap
from the band edges. On contrary to the former bifurcation, which
is exponentially small in $\epsilon$, the latter bifurcation occurs
generally in the order of ${\rm O}(\epsilon^4)$.

For the case $\mu_m = \mu_k = \mu_n$ and $\nu_{mk} = - 2
\Delta_n$, the system (\ref{couple1})--(\ref{couple2}) can be
simplified due to the obvious reduction: $\mu_m^{(2)} =
\mu_k^{(2)} = \mu_n^{(2)}$ and $\chi_{nm}^{(2)} = \chi_{nk}^{(2)}
= \chi_{nmk}^{(2)} = \chi_{nkm}^{(2)} = \chi_n^{(2)}$. Using the
variables,
\begin{equation}
   u_n = B_m + C_k, \quad w_n = i( B_m - C_k ), 
\end{equation}
and
\begin{equation}
\lambda^{(n)} = i( \Omega_{mk} - \Delta_n ),
\end{equation}
we transform the problem (\ref{couple1})--(\ref{couple2}) to the
form:
\begin{equation} \label{eigenvalue-n}
   {\cal L}_1^{(n)} u_n = - \lambda^{(n)} w_n,
   \qquad {\cal L}_0^{(n)} w_n = \lambda^{(n)} u_n,
\end{equation}
where ${\cal L}_1^{(n)}$ and ${\cal L}_0^{(n)}$ are linear
Schrodinger operators with decaying potentials:
\begin{eqnarray} \label{L0n}
   {\cal L}_0^{(n)} & = & \mu_n^{(2)} \frac{d^2}{d X^2} -
   \Delta_n + \chi_n^{(2)} A_n^2(X), \\
   \label{L1n} {\cal L}_1^{(n)} & = & \mu_n^{(2)} \frac{d^2}{d X^2} -
   \Delta_n + 3 \chi_n^{(2)} A_n^2(X),
\end{eqnarray}
where ${\rm sign}(\Delta_n) = {\rm sign}(\mu_n^{(2)}) = {\rm
sign}(\chi_n^{(2)})$. The linear eigenvalue problem
(\ref{eigenvalue-n}) is the linearized NLS problem on the real
line, associated to the sech-solitons (\ref{NLSsoliton}). The
problem has two branches of the continuous spectrum for
$\lambda^{(n)} \in i(-\infty,-|\Delta_n|) \cup
i(|\Delta_n|,\infty)$, the four-dimensional null space
$\lambda^{(n)} = 0$ and the resonance at the band edges
$\lambda^{(n)} = \pm i |\Delta_n|$. A small perturbation of the
decaying potentials in the problem (\ref{eigenvalue-n}) may result
in the edge bifurcation of a single pair of internal modes
$\lambda^{(n)} = \pm i \Omega^{(n)}$, such that $\Omega^{(n)} <
|\Delta_n|$, provided a certain integral criterion is
satisfied~\cite{Pelinovsky:1998-121:PD, Kapitula:2002-1117:SJMA}.

It was shown~\cite{Kivshar:1998-5032:PRL, Kapitula:2001-533:NLN}
that the discrete NLS equation with a small lattice step size
supports bifurcations of a single pair of internal modes from the
band edges beyond the linearized NLS problem (\ref{eigenvalue-n}).
In order to study these bifurcations, we would have to extend
perturbation series expansions (\ref{perturbation7})--(\ref{perturbation67}) 
to the next orders and derive the ${\rm O}(\epsilon^2)$ corrections 
to the linearized NLS problem (\ref{eigenvalue-n}). 
This work goes beyond the scope of the present paper. 
We only note that there is at most one pair
of internal modes bifurcating in the narrow gap near $\lambda =
0$.

\section{Conclusions} \lsect{concl}

We have presented a
systematic analysis of the existence, bifurcations, linear
stability, and internal modes of {\em gap solitons} in the
framework of the nonlinear Schr\"odinger equation with a periodic
potential. This model or its generalizations appear in a variety
of physical applications including low-dimensional photonic
crystals, arrays of coupled nonlinear optical waveguides,
optically-induced photonic lattices, and Bose-Einstein condensates
loaded onto an optical lattice. In the framework of this model, we
have classified branches of gap solitons bifurcating from the band
edges of the Floquet-Bloch spectrum, by means of the multi-scale
perturbation series expansion method. We have demonstrated that
gap solitons can appear near all lower or upper band edges of the
spectrum for focusing or defocusing nonlinearity, respectively.
For the first time to our knowledge, we have studied the
gap-soliton internal modes and stability of gap solitons in the
framework of the continuous model with a periodic potential. We
have demonstrated that the gap-soliton stability is determined by
the broken translational invariance, as well as the location of
internal modes with respect to the spectral bands of the inverted
spectrum. We have shown analytically and numerically that complex
eigenvalues of the stability problem correspond to oscillatory
instabilities of gap solitons.

The analytical results presented above are rather general, and
they are expected to be valid for different types of smooth
arbitrary-shaped periodic potentials. Although our numerical
examples have been presented for the specific case of the
sinusoidal potential, we expect that our main results can be
applied to other types of similar nonlinear models with
periodically varying parameters, such as nonlinear photonic
crystals and optically-induced photonic lattices.

\section*{Acknowledgements}

The authors acknowledge a support from the Australian Research
Council. They thank Dr. Elena Ostrovskaya for useful collaboration
and help at the initial stage of this project, as well as
F.~Gesztesy, B.~Sandstede, and A.~Scheel for useful references.
D.P. thanks the Nonlinear Physics Group for hospitality during his
visit.

\begin{widetext}
\appendix

\section{Numerical method for calculation of eigenvalues}
 \lsect{Evans}

Eigenvalues of the spectral problem~(\ref{eigenvalue})
provide a key information about the soliton stability. However,
an accurate numerical calculation of complex eigenvalues describing
oscillatory instabilities of gap solitons is a nontrivial problem
even in the case of a simpler system of coupled-mode 
equations~\cite{Barashenkov:2000-22:CPC,
Schollmann:2000-218:PA, Schollmann:2000-5830:PRE}. The reason 
for numerical difficulties is
that different components of eigenvectors have very different
localization widths. For example, the modes shown in the bottom
part of Figs.~\rpict{instabOscilSF2prof},\rpict{instabOscilSF2bprof} 
are much broader than the soliton width, while the modes 
shown in the top part of 
Figs.~\rpict{instabOscilSF2prof},\rpict{instabOscilSF2bprof} 
have comparable width. Numerical approaches used in
a number of earlier studies~\cite{Barashenkov:2000-22:CPC,
Schollmann:2000-218:PA, Schollmann:2000-5830:PRE} were based on
the discretization of Eq.~(\ref{eigenvalue}), however an accurate
description of weakly localized modes requires the use of
impractically wide computational windows. It was suggested that
the eigenvalues can be calculated approximately, and then improved
using a special iterative procedure~\cite{Schollmann:2000-218:PA,
Schollmann:2000-5830:PRE}. In our analysis, we avoid such problems
by using a different approach based on the Evans function
formalism. This method proved to be very effective
for tracing soliton instabilities in periodic
systems~\cite{Sukhorukov:2002-36609:PRE}.

First, we reformulate the spectral problem~(\ref{eigenvalue})
using a different set of functions $U = u+i w$ and $W = u - i w$,
\begin{eqnarray} \nonumber
     - \frac{d^2 U}{dx^2} + V(x) U   
    + \sigma \Phi_s^2(x) (2 U + W) & = & (\mu_s + i \lambda) U, \\
\label{UWproblem}
    - \frac{d^2 W}{dx^2} + V(x) W
      + \sigma \Phi_s^2(x) (2 W + U) & = & (\mu_s - i \lambda) W.
\end{eqnarray}
The advantage of this formulation for the numerical analysis is that
Eqs.~(\ref{UWproblem}) become uncoupled away from the soliton
core, at $|x| \rightarrow \infty$. In these regions, solutions of
Eqs.~(\ref{UWproblem}) are found in terms of Bloch functions, and
they form a natural basis for representation of solutions along the whole line,
\begin{equation} \label{UWBloch}
  \begin{array}{l} {\displaystyle
    U(x) = U_1(x) \psi_1^+ + U_2(x) \psi_2^+,
  } \\*[9pt]  {\displaystyle
    W(x) = W_1(x) \psi_1^- + W_2(x) \psi_2^-,
  } \end{array}
\end{equation}
where $\psi_{1,2}^\pm(x)$ are two linearly independent Bloch 
functions, found as solutions of Eq.~(\ref{spectrum}) 
with $\mu = \mu_s \pm i \lambda$ and $U_{1,2}$ and 
$W_{1,2}$ are unknown parameters.
By using method of variation of parameters, we set 
the constraints on $U_{1,2}$ and $W_{1,2}$:
\begin{equation} \label{dUWBloch}
  \begin{array}{l} {\displaystyle
    \frac{d U}{d x} = U_1(x) \frac{d \psi_1^+}{d x}
        + U_2(x) \frac{d \psi_2^+}{d x},
  } \\*[9pt]  {\displaystyle
    \frac{d W}{d x} = W_1(x) \frac{d \psi_1^-}{d x}
        + W_2(x) \frac{d \psi_2^-}{d x}.
  } \end{array}
\end{equation}
After substituting Eqs.~(\ref{UWBloch}),(\ref{dUWBloch}) into
Eq.~(\ref{UWproblem}), we obtain a set of first-order linear
differential equations for the amplitude functions $U_j(x)$
and $W_j(x)$, $j = 1,2$, as follows
\begin{equation} \label{UWjproblem}
  \begin{array}{l} {\displaystyle
      \frac{d U_j}{d x} = (-1)^j \sigma \Phi_s^2(x) (2 U + W)
                        \psi_{3-j}^+ / {\cal D}^+
  } \\*[9pt]  {\displaystyle
      \frac{d W_j}{d x} = (-1)^j \sigma \Phi_s^2(x) (2 W + U)
                        \psi_{3-j}^- / {\cal D}^-
  } \end{array}
\end{equation}
where the Wronskian determinants 
${\cal D}^\pm = \psi_1^\pm (d \psi_2^\pm/d x) - \psi_2^\pm
(d \psi_1^\pm /d x)$ are $x$-independent~\cite[Sect.
1.6]{Born:2002:PrinciplesOptics}. 
Whereas Eqs.~(\ref{UWjproblem}) are fully equivalent to the original eigenvalue problem, they are much better suited for numerical analysis since $U_{1,2}$ and  $W_{1,2}$ only change in the region of the soliton core, where $\Phi_s^2(x)$ is non-small.
The key advantage is that the required size of the computational window is defined by the soliton width, and does not depend on the localization of linear modes.

We seek spatially localized eigenmodes, which can exist when the Bloch functions $\psi_{1,2}^{\pm}(x)$ have complex Bloch wave-numbers $k(\mu)$, and according to Eq.~(\ref{fundamental}), one of the Bloch functions is exponentially growing whereas the other one is decaying. We assume, with no loss of generality, that $|\psi_1^{\pm}| \rightarrow 0$ at $x \rightarrow + \infty$. Then, a localized mode can form when simultaneously
\begin{equation} \label{BC1}
    \lim_{x \to +\infty} (U_2,W_2) = 0, \qquad 
    \lim_{x \to - \infty} (U_1,W_1) = 0.
\end{equation}
In order to satisfy the limits (\ref{BC1}), the following 
determinant must vanish:
\begin{equation} \label{Evans}
   {\cal E}(\lambda) = {\rm Det}\left(\begin{array}{llll}
       U_{1,u}^+(x) & U_{1,w}^+(x) & U_{1,u}^-(x) & U_{1,w}^-(x) \\
       U_{2,u}^+(x) & U_{2,w}^+(x) & U_{2,u}^-(x) & U_{2,w}^-(x) \\
       W_{1,u}^+(x) & W_{1,w}^+(x) & W_{1,u}^-(x) & W_{1,w}^-(x) \\
       W_{2,u}^+(x) & W_{2,w}^+(x) & W_{2,u}^-(x) & W_{2,w}^-(x)
   \end{array} \right)  = 0,
\end{equation}
where four particular solutions of Eqs.~(\ref{UWjproblem}) 
are introduced according to the limiting behavior:
$$
\lim_{x \to +\infty} \left( \begin{array}{cc} 
U_{1,u}^+ \\ U_{2,u}^+ \\ W_{1,u}^+ \\ W_{2,u}^+ 
\end{array} \right) = \left( \begin{array}{cc} 
1 \\ 0 \\ 0 \\ 0 \end{array} \right), \quad 
\lim_{x \to +\infty} \left( \begin{array}{cc} 
U_{1,w}^+ \\ U_{2,w}^+ \\ W_{1,w}^+ \\ W_{2,w}^+ 
\end{array} \right) = \left( \begin{array}{cc} 
0 \\ 0 \\ 1 \\ 0 \end{array} \right), \quad 
\lim_{x \to -\infty} \left( \begin{array}{cc} 
U_{1,u}^- \\ U_{2,u}^- \\ W_{1,u}^- \\ W_{2,u}^- 
\end{array} \right) = \left( \begin{array}{cc} 
0 \\ 1 \\ 0 \\ 0 \end{array} \right), \quad 
\lim_{x \to -\infty} \left( \begin{array}{cc} 
U_{1,w}^- \\ U_{2,w}^- \\ W_{1,w}^- \\ W_{2,w}^- 
\end{array} \right) = \left( \begin{array}{cc} 
0 \\ 0 \\ 0 \\ 1 \end{array} \right).
$$
Then, solution of Eqs.~(\ref{UWjproblem}) satisfying the boundary conditions~(\ref{BC1}) is found as 
$U_j = \rho_u^+ U_{j,u}^+ + \rho_w^+ U_{j,w}^+$ and 
$W_j = \rho_u^+ W_{j,u}^+ + \rho_w^+ W_{j,w}^+$, where 
$\left( \rho_u^+, \rho_w^+, \rho_u^-, \rho_w^- \right)^T$ is an eigenvector corresponding to a zero eigenvalue of the matrix in Eq.~(\ref{Evans}).

The coordinate $x$ in Eq.~(\ref{Evans}) is arbitrary, but for numerical calculations a better accuracy is achieved when it is chosen at the soliton center, $x=x_0$. The function ${\cal E}(\lambda)$ is called Evans function, and its zeros define the location of eigenvalues. We approximate zeros of ${\cal E}(\lambda)$ by finding minima $|{\cal E}(\lambda)|$ along the real axis and along the imaginary axis with a small real part, and then using a two-dimensional minimization procedure in the full complex plane.

\section{Derivation of (\ref{final-quadratic-form})} \lsect{appA}

We rewrite the derivative equation (\ref{shift})
in the equivalent form:
\begin{equation} \label{ODE-x0}
   {\cal L}_1 \frac{\partial \Phi_{\epsilon}(x;X)}{\partial x}
   + \epsilon {\cal L}_1 \frac{\partial \Phi_{\epsilon}(x;X)}{\partial X} =
   - V'(x) \Phi_{\epsilon}(x;X).
\end{equation}
Using the inner product (\ref{innerproduct}), we reduce
(\ref{ODE-x0}) to the quadratic forms:
\begin{equation} \label{quadratic-form-shift}
   \left( \frac{\partial
   \Phi_{\epsilon}}{\partial X}, {\cal L}_1 \frac{\partial
   \Phi_{\epsilon}}{\partial x} \right) + \epsilon \left(
   \frac{\partial \Phi_{\epsilon}}{\partial X}, {\cal L}_1
   \frac{\partial \Phi_{\epsilon}}{\partial X} \right) = -
   \int_{-\infty}^{\infty} V'(x) \Phi_{\epsilon}(x;X) \frac{\partial
   \Phi_{\epsilon}(x;X)}{\partial X} dx.
\end{equation}
Using the Fourier series (\ref{Fourier}), Fourier transform
(\ref{Fouriertransform}), and the expansion
(\ref{constraintsymplified}), we reduce the right-hand-side of
(\ref{quadratic-form-shift}) to the form,
\begin{eqnarray}
   - \epsilon^3 \int_{-\infty}^{\infty} V'(x) \Phi_{\epsilon}(x;X)
   \frac{\partial \Phi_{\epsilon}(x;X)}{\partial X} dx =
   \frac{\epsilon}{2} \sum_{m = -\infty}^{\infty} \frac{2 \pi i m}{d}
   \hat{W}_{n,m}\left( \frac{2 \pi m}{\epsilon d};\epsilon \right) \;
   e^{\frac{2 \pi i m x_0}{d}} & = & \frac{1}{2} M_s'(x_0).
\end{eqnarray}
On the other hand, the first term in the left-hand-side of
(\ref{quadratic-form-shift}) is identically zero:
\begin{equation} \label{zero-identically}
   \left( \frac{\partial
   \Phi_{\epsilon}}{\partial X}, {\cal L}_1 \frac{\partial
   \Phi_{\epsilon}}{\partial x} \right) = \left( {\cal L}_1
   \frac{\partial \Phi_{\epsilon}}{\partial X}, \frac{\partial
   \Phi_{\epsilon}}{\partial x} \right) = 0,
\end{equation}
which is proved from the nonlinear problem (\ref{ODE}) as follows:
\begin{eqnarray} \nonumber
   \frac{\partial \Phi_{\epsilon}}{\partial x} {\cal L}_1 \frac{\partial
   \Phi_{\epsilon}}{\partial X} & = & \frac{\partial}{\partial X}
   \left[ \frac{\partial \Phi_{\epsilon}}{\partial x} \left( -
   \Phi_{\epsilon}'' + V(x) \Phi_{\epsilon} - \mu_s \Phi_{\epsilon} +
   \sigma \epsilon^2 \Phi_{\epsilon}^3 \right) \right] \\
   & - & \frac{\partial^2 \Phi_{\epsilon}}{\partial X \partial x}
   \left( - \Phi_{\epsilon}'' + V(x) \Phi_{\epsilon} - \mu_s
   \Phi_{\epsilon} + \sigma \epsilon^2 \Phi_{\epsilon}^3 \right) = 0.
\end{eqnarray}
The quadratic form in (\ref{quadratic-form-shift}) is therefore
rewritten in the final form (\ref{final-quadratic-form}).

\section{Derivation of (\ref{final-Fermi-rule})} \lsect{appB}

We extend the perturbation series expansions (\ref{perturbation4a}),
(\ref{perturbation4b}), and (\ref{perturbation4}) to the higher
orders in powers of $\epsilon$. Solving the system
(\ref{nonhomgen1}) and (\ref{nonhomgen2}) for the second-order
correction terms $(u_m^{(2)},w_m^{(2)})$, we write the solution in
the implicit form,
\begin{eqnarray} \label{nonhomgen-solution1}
   u^{(2)}_m = B_m''(X) \psi_m^{(2)}(x) +
   2 A_n^2(X) B_m(X) \psi_{nm}^{(nl2)}(x) + A_n^2(X) B_m(X) \phi_{nm}^{(nl 2)}(x), \\
   \label{nonhomgen-solution2} w^{(2)}_m = B_m''(X) \psi_m^{(2)}(x) +
   2 A_n^2(X) B_m(X) \psi_{nm}^{(nl2)}(x) - A_n^2(X) B_m(X)
   \phi_{nm}^{(nl2)}(x),
\end{eqnarray}
where $\psi_m^{(2)}(x)$ is defined in (\ref{nonhomogeneous1}),
while $\psi_{nm}^{(nl2)}(x)$ and $\phi_{nm}^{(nl2)}(x)$
solve the non-homogeneous problems,
\begin{eqnarray} \label{nonhomogeneous6}
   - \left( \psi_{nm}^{(nl2)} \right)'' +
   V(x) \psi_{nm}^{(nl2)} - \mu_m \psi_{nm}^{(nl2)} & = &
   \chi_{nm}^{(2)} \psi_m -
   \sigma \psi_n^2 \psi_m, \\
   \label{nonhomogeneous7}
   - \left( \phi_{nm}^{(nl2)} \right)'' + V(x) \phi_{nm}^{(nl2)}
   + (\mu_m -2 \mu_s) \phi_{nm}^{(nl2)} & = & - \sigma \psi_n^2 \psi_m.
\end{eqnarray}
The first equation (\ref{nonhomogeneous6}) defines periodic or
anti-periodic functions $\psi_{nm}^{(nl2)}(x)$, since the Fredholm
condition is satisfied by the relation (\ref{chinm2}). The second
equation (\ref{nonhomogeneous7}) defines a non-periodic function
$\phi_{nm}^{(nl2)}(x)$, since $\mu_r \in \Sigma_{\rm band}$, where
$\mu_r = 2 \mu_s - \mu_m$. We use the Sommerfeld radiation
boundary conditions for function $\phi_{nm}^{(nl2)}(x)$ in the
ends of the period $x \in [0,d]$:
\begin{equation}
\label{Sommerfeld} \left( \phi_{nm}^{(nl2)} \right)'(d) - i k_r
\phi_{nm}^{(nl2)}(d) = 0, \qquad \left( \phi_{nm}^{(nl2)}
\right)'(0) + i k_r \phi_{nm}^{(nl2)}(0) = 0,
\end{equation}
where $k_r = k(\mu_r)$ and the dispersion relation $k(\mu)$ is
defined in the Bloch functions (\ref{fundamental}). The Sommerfeld
boundary conditions (\ref{Sommerfeld}) imply that the
time-dependent solution $\Psi(x,t)$ of the NLS equation
(\ref{NLS}) linearized at the gap soliton $\Phi_s(x) e^{-i \mu_s
t}$ takes the form of outgoing radiative waves (\ref{fundamental})
directed outwards the period $x \in [0,d]$:
\begin{equation} \label{outgoing-wave}
   \Psi(x,t) - \Phi_s(x) e^{- i \mu_s t}
   \rightarrow \left\{ \begin{array}{cc} \alpha_+(X) \phi_1(x) e^{i k_r x
   - i \mu_r t}, \quad x \to d^- \\ \alpha_-(X) \phi_2(x) e^{-i k_r x - i
   \mu_r t}, \quad x \to 0^+,
   \end{array} \right.
\end{equation}
where $\alpha_{\pm}(X)$ are amplitudes of the radiative waves. The
Sommerfeld boundary conditions were used recently for embedded
solitons~\cite{Pelinovsky:2002-1469:PRSA}. Since the Sommerfeld
boundary conditions (\ref{Sommerfeld}) are not symmetric, the
functions $\phi_{nm}^{(nl2)}(x)$ are complex-valued. The
complex-valued functions $\phi_{nm}^{(nl2)}(x)$ result in
complex-valued corrections to imaginary eigenvalues $\lambda$ in
higher orders of the perturbation series expansion
(\ref{perturbation4}). In order to avoid lengthy analysis of the
perturbation series equations at the third and fourth orders of
$\epsilon$ and to capture the non-zero real part of complex
eigenvalues $\lambda$, we rewrite the linearized stability problem
(\ref{eigenvalue}) in the form of the balance equation:
\begin{equation} \label{quadratic-form-system}
   \left( \lambda + \bar{\lambda} \right)
   \left( u \bar{w} - \bar{u} w \right) = \frac{d}{dx}
   \left( u \frac{d \bar{u}}{dx} - \frac{du}{dx} \bar{u}
   + w \frac{d \bar{w}}{dx} - \frac{dw}{dx} \bar{w} \right).
\end{equation}
Using perturbation series expansions (\ref{perturbation4a}),
(\ref{perturbation4b}), and (\ref{perturbation4}), we rewrite
(\ref{quadratic-form-system}) in variables $x$ and $X$. The first
non-zero term occurs at the fourth order of $\epsilon$ and takes
the form:
\begin{equation} \label{Fermi1}
   -4i {\rm Re}(\lambda) B_m^2(X) \psi_m^2(x) =
   \epsilon^4 \frac{\partial Q_4(x;X)}{\partial x} + {\rm
   O}(\epsilon^5),
\end{equation}
where the fourth-order correction term $Q_4(x;X) =
Q_4^{(per)}(x;X) + Q_4^{(np)}(x;X)$ is decomposed in a periodic
function of $x$ and a non-periodic function of $x$, the latter is
given by
\begin{equation} \label{Q4}
   Q_4^{(nl)}(x;X) = u_m^{(2)} \left(
   \bar{u}_m^{(2)}\right)' - \bar{u}_m^{(2)} \left( u_m^{(2)}
   \right)' + w_m^{(2)} \left( \bar{w}_m^{(2)} \right)' -
   \bar{w}_m^{(2)} \left( w_m^{(2)} \right)'.
\end{equation}
The prime in (\ref{Q4}) denotes the derivative in $x$. Integrating
(\ref{Fermi1}) over the period $x \in [0,d]$ and over the real
line of $X$, and using the explicit representation
(\ref{nonhomgen-solution1}) and (\ref{nonhomgen-solution2}) for
$u_m^{(2)}(x;X)$ and $w_m^{(2)}(x;X)$, we rewrite the balance
equations (\ref{Fermi1}) and (\ref{Q4}) in the form:
\begin{eqnarray} \nonumber
   -4i {\rm Re}(\lambda) \left( \int_{-\infty}^{\infty}
   B_m^2 dX \right) \left( \int_0^d \psi_m^2 dx \right) = \\
   \label{Fermi2}
   2 \epsilon^4 \left( \int_{-\infty}^{\infty} A_n^4 B_m^2 dX \right)
   \left( \phi_{nm}^{(nl2)} \left( \bar{\phi}_{nm}^{(nl2)} \right)' -
   \bar{\phi}_{nm}^{(nl2)} \left( \phi_{nm}^{(nl2)} \right)' \right)
   \biggr|_{x = 0}^{x = d} + {\rm O}(\epsilon^5).
\end{eqnarray}
Using the Sommerfeld boundary conditions (\ref{Sommerfeld}), we
finally derive the expansion (\ref{final-Fermi-rule}), where
\begin{equation} \label{Fermi3}
   \Gamma_m = \left( \frac{\int_{-\infty}^{\infty}
   A_n^4 B_m^2 dX}{ \int_{-\infty}^{\infty} B_m^2 dX} \right) \left(
   \frac{k_r \left( |\phi_{nm}^{(nl2)}(0)|^2 +
   |\phi_{nm}^{(nl2)}(d)|^2 \right)}{\int_0^d \psi_m^2 dx} \right)
   \geq 0.
\end{equation}
The formula (\ref{Fermi3}) is referred to
as the Fermi golden rule of radiative decay of embedded
eigenvalues~\cite{Tsai:2002-1629:IMRN, Pelinovsky:2002-1469:PRSA}.
\end{widetext}

\end{sloppy}
\end{document}